\documentclass[twoside,twocolumn,9pt]{article}
\usepackage{extsizes}
\usepackage[super,sort&compress,comma]{natbib} 
\usepackage[version=3]{mhchem}
\usepackage[left=1.5cm, right=1.5cm, top=1.785cm, bottom=2.0cm]{geometry}
\usepackage{balance}
\usepackage{widetext}
\usepackage{times,mathptmx}
\usepackage{sectsty}
\usepackage{amsmath}
\usepackage{amssymb}
\usepackage{mwe}
\usepackage{soul}
\usepackage{MnSymbol}
\usepackage{graphicx} 
\usepackage{lastpage}
\usepackage[format=plain,justification=justified,singlelinecheck=false,font={stretch=1.125,small,sf},labelfont=bf,labelsep=space]{caption}
\usepackage{float}
\usepackage{fancyhdr}
\usepackage{fnpos}
\usepackage[english]{babel}
\usepackage{array}
\usepackage[inline]{enumitem}
\usepackage{droidsans}
\usepackage{charter}
\usepackage[T1]{fontenc}
\usepackage[usenames,dvipsnames]{xcolor}
\usepackage{setspace}
\usepackage[compact]{titlesec}

\usepackage{bm}
\usepackage{epstopdf}

\definecolor{cream}{RGB}{222,217,201}

\begin{document}

\pagestyle{fancy}
\thispagestyle{plain}
\fancypagestyle{plain}{

\fancyhead[C]{\includegraphics[width=18.5cm]{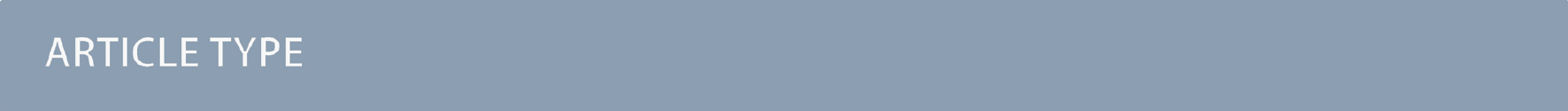}}
\fancyhead[L]{\hspace{0cm}\vspace{1.5cm}\includegraphics[height=30pt]{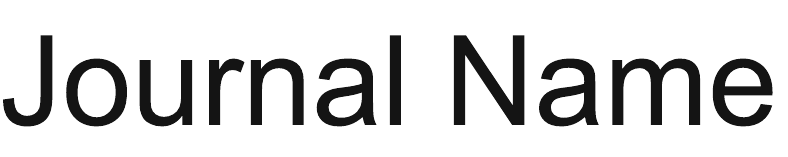}}
\fancyhead[R]{\hspace{0cm}\vspace{1.7cm}\includegraphics[height=55pt]{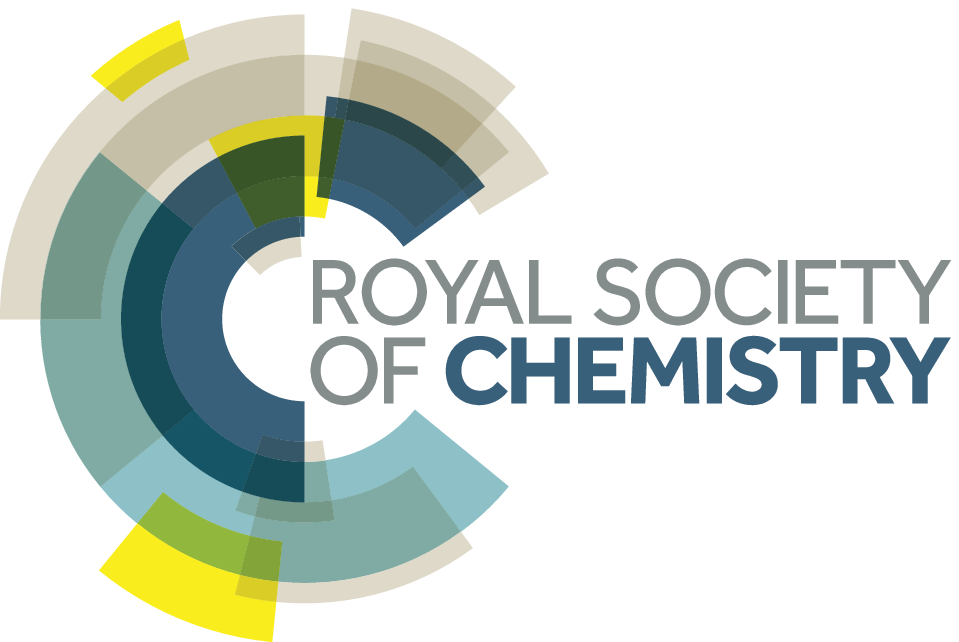}}
\renewcommand{\headrulewidth}{0pt}
}

\makeFNbottom
\makeatletter
\renewcommand\LARGE{\@setfontsize\LARGE{15pt}{17}}
\renewcommand\Large{\@setfontsize\Large{12pt}{14}}
\renewcommand\large{\@setfontsize\large{10pt}{12}}
\renewcommand\footnotesize{\@setfontsize\footnotesize{7pt}{10}}
\makeatother

\renewcommand{\thefootnote}{\fnsymbol{footnote}}
\renewcommand\footnoterule{\vspace*{1pt}%
\color{cream}\hrule width 3.5in height 0.4pt \color{black}\vspace*{5pt}} 
\setcounter{secnumdepth}{5}

\makeatletter 
\renewcommand\@biblabel[1]{#1}            
\renewcommand\@makefntext[1]%
{\noindent\makebox[0pt][r]{\@thefnmark\,}#1}
\makeatother 
\renewcommand{\figurename}{\small{Fig.}~}
\sectionfont{\sffamily\Large}
\subsectionfont{\normalsize}
\subsubsectionfont{\bf}
\setstretch{1.125} 
\setlength{\skip\footins}{0.8cm}
\setlength{\footnotesep}{0.25cm}
\setlength{\jot}{10pt}
\titlespacing*{\section}{0pt}{4pt}{4pt}
\titlespacing*{\subsection}{0pt}{15pt}{1pt}

\fancyfoot{}
\fancyfoot[LO,RE]{\vspace{-7.1pt}\includegraphics[height=9pt]{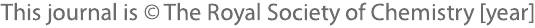}}
\fancyfoot[CO]{\vspace{-7.1pt}\hspace{13.2cm}\includegraphics{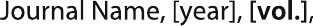}}
\fancyfoot[CE]{\vspace{-7.2pt}\hspace{-14.2cm}\includegraphics{head_foot/RF}}
\fancyfoot[RO]{\footnotesize{\sffamily{1--\pageref{LastPage} ~\textbar  \hspace{2pt}\thepage}}}
\fancyfoot[LE]{\footnotesize{\sffamily{\thepage~\textbar\hspace{3.45cm} 1--\pageref{LastPage}}}}
\fancyhead{}
\renewcommand{\headrulewidth}{0pt} 
\renewcommand{\footrulewidth}{0pt}
\setlength{\arrayrulewidth}{1pt}
\setlength{\columnsep}{6.5mm}
\setlength\bibsep{1pt}

\makeatletter 
\newlength{\figrulesep} 
\setlength{\figrulesep}{0.5\textfloatsep} 

\newcommand{\topfigrule}{\vspace*{-1pt}%
\noindent{\color{cream}\rule[-\figrulesep]{\columnwidth}{1.5pt}} }

\newcommand{\botfigrule}{\vspace*{-2pt}%
\noindent{\color{cream}\rule[\figrulesep]{\columnwidth}{1.5pt}} }

\newcommand{\dblfigrule}{\vspace*{-1pt}%
\noindent{\color{cream}\rule[-\figrulesep]{\textwidth}{1.5pt}} }

\makeatother

\twocolumn[
  \begin{@twocolumnfalse}
\vspace{3cm}
\sffamily
\begin{tabular}{m{4.5cm} p{13.5cm} }

\includegraphics{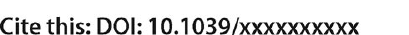} & \noindent\LARGE{\textbf{Self-assembly and rheology of dipolar colloids in simple shear - studied by multi-particle collision dynamics}} \\
\vspace{0.3cm} & \vspace{0.3cm} \\

 & \noindent\large{Dmitry Zablotsky,\textit{$^{a,b}$} Elmars Blums,\textit{$^{b}$} and Hans J. Herrmann\textit{$^{a}$}} \\

\includegraphics{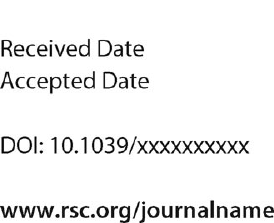} & \noindent\normalsize{Magnetic nanoparticles in a colloidal solution self-assemble in various aligned structures, which has a profound influence on the flow behavior. However, the precise role of the microstructure in the development of the rheological response has not been reliably quantified. We investigate the self-assembly of dipolar colloids in simple shear using hybrid molecular dynamics and multi-particle collision dynamics simulations with explicit coarse-grained hydrodynamics; conduct simulated rheometric studies and apply micromechanical models to produce master curves, showing evidence of the universality of the structural behavior governed by the competition of the bonding (dipolar) and erosive (thermal and/or hydrodynamic) stresses. The simulations display viscosity changes across several orders of magnitude in fair quantitative agreement with various literature sources, substantiating the universality of the approach, which seems to apply generally across vastly different length scales and a broad range of physical systems.

} \\

\end{tabular}

 \end{@twocolumnfalse} \vspace{0.6cm}

  ]

\renewcommand*\rmdefault{bch}\normalfont\upshape
\rmfamily
\section*{}
\vspace{-1cm}


\footnotetext{\textit{$^{a}$~ETH Zurich, Computational Physics for Engineering Materials, Institute for Building Materials, Stefano-Franscini-Platz 3, CH-8093 Zurich, Switzerland.}}
\footnotetext{\textit{$^{b}$~University of Latvia, Raina bulvaris 19, LV-1586 Riga, Latvia. E-mail: dmitrijs.zablockis@lu.lv}}


\section{Introduction}

Self-assembly is a pathway to novel materials\cite{Nie_Petukhova_Kumacheva_2009} with structural hierarchy, multi-functionality and programmed response to mechanical stress or external field\cite{Grzelczak_Vermant_Furst_Liz-Marzan_2010}.
In magnetic colloids - ferrofluids - the magnetic nanoparticles are suspended in a non-magnetic liquid carrier and contain a spontaneously polarized monodomain \citep{blums1997magnetic}. The ability of the ferrofluids to actively interact with a magnetic field, while retaining fluidity is the basis of their practical utility in various technological devices\cite{Raj_Moskowitz_1990} and novel applications in magnetic chemistry\cite{Polshettiwar_Luque_Fihri_Zhu_Bouhrara_Basset_2011} and medicine\cite{SUN_LEE_ZHANG_2008}. 
The suspensions of magnetic particles\cite{deVicente_Klingenberg_Hidalgo-Alvarez_2011} exhibit enhanced viscoelasticity as so far as to display a rapid, tractable and reversible field-induced rheological transition between a mostly Newtonian liquid and a plastic solid, enabling magnetomechanical devices and technologies, which take advantage of the tunable flow properties and can be continuously adjusted for improved performance in automotive\citep{Carlson_Catanzarite_StClair_1996}, mechatronic or biomechatronic/prosthetic applications \citep{Klingenberg_2001} or for protection of civil constructions, machine tools and industrial installations against vibrations or seismic shock \citep{Xu_Qu_Ko_2000,Chen_Wang_Ko_Ni_SpencerJr_Yang_2003}.
The electrical analogs, i.e. ferroelectric suspensions of highly polarizable particles in a nonpolarizable solvent, also display an appreciable enhancement of flow resistance \citep{Wen_Huang_Yang_Lu_Sheng_2003} under electric field.
Magnetic colloids are adaptable platforms for composite soft magnetic materials with customized physical properties \citep{Galindo-Gonzalez_deVicente_Ramos-Tejada_Lopez-Lopez_Gonzalez-Caballero_Duran_2005,Galicia_Cousin_Dubois_Sandre_Cabuil_Perzynski_2009,Santiago-Quinonez_Rinaldi_2012} also exhibiting viscoelastic behavior\cite{de_Gans_Duin_van_den_Ende_Mellema_2000,Volkova_Bossis_Guyot_Bashtovoi_Reks_2000}. 
The new generation of magnetic colloids based on nanoparticles with accurately tailored size and/or oxide-free magnetic materials feature greatly enhanced interaction strength \citep{Hess_Parker_1966, Sun_Murray_1999, Butter_Bomans_Frederik_Vroege_Philipse_2003b, Klokkenburg_Dullens_Kegel_Erne_Philipse_2006, Lopez-Lopez_Gomez-Ramirez_Rodriguez-Arco_Duran_Iskakova_Zubarev_2012}. 
The anisotropic bonding of the polarizable particles or spontaneously polarized nanoparticles and their directional assembly \citep{Grzelczak_Vermant_Furst_Liz-Marzan_2010} under exposure to the external electromagnetic fields induces the thickening of the dipolar materials subjected to a macroscopic flow\cite{Lopez-Lopez_Gomez-Ramirez_Rodriguez-Arco_Duran_Iskakova_Zubarev_2012,Wen_Huang_Yang_Lu_Sheng_2003}. The associated nonlinear rheology is sensitive to the flow-induced conformational changes of the dissolved microstructure.
The safe and effective application of these systems in diverse areas, i.a. related to human health, relies on understanding and quantifying their microstructural and rheological properties. 
The characterization of the mechanisms providing the rheological specificity facilitates the development of more efficient materials and technological devices and processes, where their design extensively relies on phenomenology or structural rheology\cite{Berli_de_Vicente_2012}. One of the difficulties is the absence of a rigorous micromechanical model for the associated micro-macro structure-property relationships. The involved self-assembly scenarios are difficult to quantify and despite a large amount of experimental data there is only limited reliable theoretical understanding beyond simpler models confined to the vicinity of the equilibrium state. Direct methods for correlating and interpreting experimental measurements across vastly different systems would deliver substantial advance to the field. In turn, for suspensions of permanent dipoles, e.g. ferromagnetic (or ferroelectric) colloids, where thermal fluctuations play a significant role, comparatively little is known about the structural transitions and conformational changes introduced by the competition of the orientational influence of the shear flow and an external electromagnetic field. The usual dimensional scaling is not applicable in this regime and the interpretation of experimental data is particularly challenging.
In this paper we consider the microstructural transformations and the associated rheological response of sheared suspensions of thermalized spherical particles with embedded point dipoles as a model for dipolar colloids in an external field. We verify several developed micromechanical and micro-macro (rheological) structure-property relationships using the controlled environment of particle-level simulations to resolve the interplay between the anisotropic pair interactions, thermal fluctuations, solvent-mediated hydrodynamics and external fields. We employ a theoretical analysis and perform a validation against available experimental data to access a wide range of relevant scales and physical parameters for dipolar systems of substantially different physical nature to demonstrate the universality of the approach.

The paper is organized as follows: In Section~\ref{sec:model} we describe our model and the simulation approach. We exploit a hybrid MD-MPCD method implementing coarse-grained hydrodynamics. In Section~\ref{sec:results} we conduct simultaneous measurements of the micro (structural) and macro (rheological) characteristics: we start by revisiting some features of the equilibrium state (Section~\ref{sec:equilibrium}). In Section~\ref{sec:non-equilibrium} we consider the conformational transformations and the erosion of microstructure in a simple shear flow. The viscometric properties are reported in Section~\ref{sec:rheological_properties} and the comparison with experiments is discussed in Section~\ref{sec:experiments} in terms of the master curves. Finally, in Section~\ref{sec:phase_diagram} we consider the shear-induced mesophase transformations. To conclude, Section~\ref{sec:conclusions} provides the summary of this work.

\section{Model}
\label{sec:model}
Non-equilibrium molecular dynamics simulations provide a direct quantitative link between the microstructure and the macroscopic properties. Free-draining Brownian dynamics has been used extensively in the past 
\cite{Whittle_1990, Klingenberg_van_Swol_Zukoski_1989, Melrose_1992, Martin_2000}, 
however, the idealized oversimplified hydrodynamics yields poor quantitative agreement with the experiments, underestimating the rheological properties by at least an order of magnitude \citep{Whittle_1990, Klingenberg_van_Swol_Zukoski_1991a, Klingenberg_van_Swol_Zukoski_1991b, Joung_See_2008}. Stokesian dynamics \citep{Brady_Bossis_1988} includes a more rigorous treatment of the particle-solvent coupling and many body hydrodynamics, but is restricted to very small systems and/or reduced dimensionality: monolayer simulations\citep{Baxter_Drayton_Brady_1996} (25 particles) of the ER effect reported a fair quantitative agreement with measured rheometric data and confirmed the relative importance of the hydrodynamic contributions to the total suspension stress. A lattice Boltzmann study \citep{Kim_Stratford_Camp_Cates_2009} has shown that the role of the hydrodynamic interactions remains modest ($\phi=0.03-0.20$ and $\lambda=4, 8$) in unsheared ferrofluids. In this work we use multi-particle collision dynamics\cite{Malevanets_Kapral_1999} employing a stochastic bath of point-particles for coarse-graining the solvent and thermalization. This method also resolves the hydrodynamic modes on the desired length scale and is computationally efficient.

\subsection{Dimensionless parameters}
\label{sec:dimensionless_parameters}
The structural and rheological complexity of sheared dipolar colloids is sufficiently described in terms of the volume fraction $\phi$, the dipole-dipole interaction parameter $\lambda$ and the Mason number $Mn$. The interaction parameter $\lambda$ characterizes the ability of the colloid to form an equilibrium structure and expresses the ratio of the characteristic energy of the dipole-dipole interaction between the particles to the average energy of their thermal fluctuations:
\begin{equation}
  \lambda = \frac{1}{T^{*}} = \frac{\mu_{0}}{4\pi}\frac{\mu^{2}}{\sigma^{3}kT}
\end{equation}
where $\mu_{0}$ is the magnetic constant, $\mu$ - magnitude of the particle dipole moment, $\sigma$ - distance of closest proximity, i.e. the effective diameter of the particles, including the thickness of the solvation shell. The Mason number\citep{Chaffey_Mason_1968} is a ratio of the erosive hydrodynamic forces to the bonding polarization forces encompassing the aggregate stress and carries the meaning of a reduced shear rate
\begin{equation}
  Mn = \frac{\pi\eta_{0}\sigma^{3}}{2\lambda kT}\dot{\gamma}\label{eq:mason}
\end{equation}
where $\eta_{0}$ is the viscosity of the solvent and $\dot{\gamma}$ is the characteristic shear rate. The Mason number and $\lambda$ are related via the Peclet number $Mn=\frac{2}{3\lambda}Pe$. The temperature only enters the $\lambda$ parameter - for nanometric colloids and finite $\lambda$ the thermal effects remain an important factor.

\subsection{Simulation method}
\label{sec:simulation_method}
The dipolar colloids are modeled as a monodisperse solution of spherical grains with a diameter $\sigma$ and an embedded point dipole $\bm{\mu}$. The hybrid approach \cite{Hecht_Harting_Ihle_Herrmann_2005} combines the molecular dynamics (MD) simulation of the interacting colloids immersed in a coarse-grained fluid governed by the multi-particle collision dynamics (MPCD) \cite{Malevanets_Kapral_1999,Malevanets_Kapral_2000}.
The MD-MPCD\cite{Kapral,Gompper_Ihle_Kroll_Winkler}, introduced by Malevanets and Kapral, is an established approach to study particle suspensions \cite{Padding_Louis_2004, Padding_Louis_2008}, instabilities \cite{Wysocki_Royall_Winkler_Gompper_Tanaka_vanBlaaderen_Lowen_2009, Wysocki_Lowen_2011}, colloids with isotropic attractions \cite{Hecht_Harting_Ihle_Herrmann_2005, Moncho-Jorda_Louis_Padding_2010}, anisometric colloids \cite{Ripoll_Winkler_Mussawisade_Gompper_2008}, polymer-like models \cite{Malevanets_Yeomans_2000, Padding_Briels_2003, Fedosov_Singh_Chatterji_Winkler_Gompper_2012,  Myung_Taslimi_Winkler_Gompper_2014}; polyelectrolytes\cite{Hickey_Shendruk_Harden_Slater_2012}, membranes \cite{Noguchi_Gompper_2004}, active matter \cite{Tao_Kapral_2009, Zottl_Stark_2012, Schaar_Zottl_Stark_2015} and self-assembly scenarios \cite{Huang_Kapral_Mikhailov_Chen_2012}. For the first time this method is applied here to a system with anisotropic (dipolar\cite{Weis_Levesque}) interactions
\begin{equation}
  \beta U_{dd}\left(\bm{\mu_{i}},\bm{\mu_{j}},\bm{r_{ij}}\right)=-\lambda \left[ \frac{3 \left(\bm{\mu_{i}}\cdot\bm{r_{ij}}\right)\left(\bm{\mu_{j}}\cdot\bm{r_{ij}}\right)}{r_{ij}^{5}} - \frac{\bm{\mu_{i}}\cdot\bm{\mu_{j}}}{r_{ij}^{3}} \right] \label{eq:DD}
\end{equation}
The excluded volume interactions between the colloids, incl. the steric hindrance of the solvation shell, are modeled by the repulsive Weeks-Chandler-Andersen (WCA) potential ($n=6$), which is a combination of soft potentials
\begin{equation}
  \beta U_{WCA}\left(r_{ij}\right)=4 \left[\left(r_{ij}\right)^{-2n}-\left(r_{ij}\right)^{-n}+\frac{1}{4}\right],\quad r_{ij}\leq \sqrt[n]{2} \label{eq:WCA}
\end{equation}
where $\bm{r_{ij}}$ is the center-to-center distance between particles $i$ and $j$, $\beta=\left(k T\right)^{-1}$. All lengths are given in units of particle diameter $\sigma$. 

In the MPCD approach the coarse-grained solvent is implemented by an ensemble of point particles with mass $m_{f}$. Their dynamics proceeds in alternating streaming and collision steps. During the streaming the particles propagate ballistically during $\Delta t_{mpcd}$
\begin{equation}
\bm{r}_i\left(t + \Delta t_{mpcd}\right) = \bm{r}_i\left(t\right) + \Delta t_{mpcd}\bm{v}_i\left(t\right)
\end{equation} 
For the subsequent collision step the simulated volume is equipartitioned and the particles are assigned to cubic cells (bins) $a\times a\times a$, which define the effective collision environment. The multiparticle collisions are modeled by collective stochastic rotations, whereby the peculiar (thermal) velocities are transformed on a bin-wise basis
\begin{equation}
\bm{v}_i \rightarrow \left\langle\bm{v}\right\rangle_{J}+\kappa_J\bm{\Omega}_{J}\left(\alpha\right)\cdot\left(\bm{v}_i-\left\langle\bm{v}\right\rangle_{J}\right)
\end{equation} 
where $\bm{v}_i$ is the velocity of i\textsuperscript{th} MPCD-particle in the J\textsuperscript{th} bin containing $n_J$ particles, i.e. $i\in\left\lbrace 1,\dots, n_J\right\rbrace$; $\left\langle\bm{v}\right\rangle_{J}=n_J^{-1}\sum_{i=1}^{n_J}{\bm{v}_i}$ is the velocity bias and $ \bm{\Omega}_{J}\left(\alpha\right)$ is the matrix for a rotation by an angle $\alpha=\frac{\pi}{2}$ around one of the six orthogonal axes aligned with the coordinate axes of the simulation box\cite{Malevanets_Kapral_1999, Tuzel_Strauss_Ihle_Kroll_2003, Hecht_Harting_Ihle_Herrmann_2005} randomly selected for every bin and collision step. The local scaling factor $\kappa_J$
is used to maintain a constant temperature $T$ in non-equilibrium simulations. The bin-wise conservation of the streaming momentum by the coarse-grained collision scheme leads to correct hydrodynamics on the length scales $r>a$. The stochastic multi-particle exchanges induce the viscosity of the MPCD fluid\cite{Ihle_Tuzel_Kroll_2004}.
The colloid-solvent coupling is achieved at the streaming step \cite{Hecht_Harting_Ihle_Herrmann_2005} using the stochastic BCs \cite{Inoue2002, Hecht_Harting_Ihle_Herrmann_2005} to enforce the no-slip condition. 
The momentum exchanges are calculated explicitly\cite{Hecht_Harting_Ihle_Herrmann_2005}, imparting force and torque on the dissolved solutes. Hence, the colloids are correctly thermalized and the translational and rotational Brownian motion is recovered, because momentum conservation is ensured at each collision. 
The choice of the simulation parameters preserves the hierarchy of the physical and simulation timescales as described previously \cite{Hecht_Harting_Ihle_Herrmann_2005, Padding_Louis_2006} to ensure consistency with typical colloidal systems: the ratio of the mass of the colloid $M_p$ to the mass of the MPCD particle $m_f$ is $79$, the  nominal number of the MPCD particles per bin is $\left\langle n_J \right\rangle=6$. The extent of the collision environment $a$ defines the spatial accuracy of the hydrodynamic flow\cite{Padding_Louis_2006} and is set to $a=0.5\sigma$ to properly resolve the flow on the scale of the aggregates $n\geq 2$, whose contribution to the viscosity scales as $\propto n^3$. The mean free path of the MPCD particles is $l_{mpcd}=\Delta t_{mpcd}\left(\sqrt{\beta m_f}\right)^{-1}=0.3\sigma$ ensuring Galilean invariance\cite{Ihle_Tuzel_Kroll_2004,Hecht_Harting_Ihle_Herrmann_2005}. The time between the multi-particle collisions is $\Delta t_{mpcd}=0.034\tilde{t}$, where $\tilde{t}=\sigma\sqrt{\beta M_p}$.
The simulations are performed with LAMMPS\cite{Plimpton_1995, Petersen10} in a rheometric slit channel of extent $L_x$, $L_y$, $L_z$ (volume $V=L_xL_yL_z$). For zero-field simulations $L_x=L_y=L_z=24\sigma$, whereas, for simulations with an applied field (in $y$ direction) the size of the box is increased for larger interaction parameters $\lambda$: $L_x=L_z=16\sigma$ ($\lambda<4$) or $24\sigma$ ($\lambda\geq 4$) and $L_y=32\sigma$ ($\lambda<4$), $48\sigma$ ($\lambda=4$), $72\sigma$ ($\lambda\geq 5$) to accommodate the growing microstructure. Periodic BCs are used in the $x$ (flow) and $z$ (vorticity) directions, in the $y$ (gradient) direction the channel is bounded by the free-slip walls realizing reflections $v_n\rightarrow -v_n$ of the incoming MPCD particles and a short-range repulsive potential of the form \eqref{eq:WCA} is applied to the approaching colloids to prevent percolation of the microstructure at larger $\lambda$.
In the flow simulations the shearing is induced by adjusting the streaming velocities $\left\langle\bm{v}\right\rangle_{J}$ of the MPCD particles in the bins nearest to the channel walls.
Bin-wise sampling was used to measure the flow profile and determine the actual shear rate $\dot{\gamma}$. Every simulation was repeated between $4$ and $32$ times to produce statistically independent trajectories and reduce scatter in the measured observables. The error is reported in terms of the standard deviation.

In the following we consider the two limiting situations: 
\begin{enumerate*}[label=(\alph*)]
\item strong external fields $B\rightarrow\infty$ and 
\item zero external field $B=0$ with permanent dipoles 
\end{enumerate*}. 
The strength of the dipolar interactions is varied by adjusting the $\lambda$ parameter. 

A proximity-based clustering criterion is used to identify and quantify the self-assembled microstructure: the particles are considered linked if their center-to-center distance does not exceed $r_{c}=1.1\sigma$. Other definitions\cite{Weis_Levesque_1993, Stevens_Grest_1995a} can be employed to target the specific structural elements, e.g. linear chains. The strong anisotropy of the dipolar interaction implies that any two particles can only approach in a head-to-tail configuration forming linear structures (chains, rings or branched clusters), hence, $r_{c}$ corresponds to the attractive energy $-1.5\lambda kT$, i.e. $\frac{3}{4}$ of the equilibrium bond energy, consistent with the previous studies \cite{Stevens_Grest_1995a}. 


\section{Results}
\label{sec:results}
\subsection{Conformational properties}
\label{sec:conformational_properties}

To characterize the conformational and statistical properties of the self-assembled microstructure the function $P_n\left(\bm{r}\right)$ is introduced, which is a positional probability density distribution function (PDF) due to the fact that the quantity $P_n\left(\bm{r}\right)dV$ expresses the probability of finding a colloidal particle (monomer) in the vicinity of point $\bm{r}$ relative to the center of mass of an identified aggregate ($n$-mer). It obeys a normalization condition 
\begin{equation}
\sum_{n=1}^{\infty}\int_V P_n\left(\bm{r}\right)dV = 1
\end{equation}
Figure~\ref{fgr:zero_snapshots} shows the time-averaged aggregate PDF $P\left(\bm{r}\right)=\sum_{n=2}^{\infty}P_n\left(\bm{r}\right)$ in equilibrium at zero applied field (Note: single monomer contributions are substracted here for better visualization) representing
compact globular clusters, which initially comprise relatively narrow shells of probability corresponding to the orbits of the monomers around the center of mass of the aggregates; and become rather diffuse with outward distance due to the flexibility of the fluctuating chains. 
The $n$-mer size (chain length) distribution $g_n$ representing the number of $n$-mers per unit volume of the colloid under the specific conditions of colloid concentration $\phi$, dipolar interaction strength $\lambda$ or external field is obtainable from the PDF:
\begin{equation}
  g_n=\frac{\phi}{n v_p}\int_V P_n\left(\bm{r}\right)dV \label{eq:gn_prob}
\end{equation}
where $v_p=\frac{1}{6}\pi\sigma^3$ is the volume of a colloidal particle.
The distribution $g_n$ determines the statistical and rheological properties of the ensemble of dipolar particles.

\subsubsection{Equilibrium}
\label{sec:equilibrium}

\begin{figure}[!h]
\centering
    \includegraphics[width=0.495\linewidth]{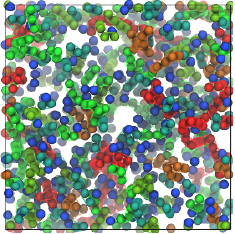}
    \includegraphics[width=0.495\linewidth]{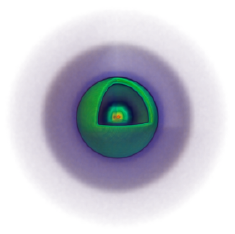}
  \caption{Left: simulation snapshot (box size $24\sigma\times 24\sigma\times 24\sigma$) of spontaneous association of dipolar particles without an applied field at a volume fraction $\phi=5\%$ and dipole interaction parameter $\lambda=4$, particles color-coded by chain length $n$: monomers ($n=1$) are blue, $n\geq10$ - red. Right: corresponding aggregate PDF $P\left(\bm{r}\right)$ of the particles in the clusters (green shell represents the orbit of dimers).
}
  \label{fgr:zero_snapshots}
\end{figure}

The onset of chaining can be estimated from the second $\sim \phi^2 $ virial coefficient $b_2$\citep{Gennes_Pincus_1970} of the dipolar colloids in zero $-\frac{e^{2\lambda}}{3\lambda^3}$ and strong field $-\frac{e^{2\lambda}}{3\lambda^2}$. From $|b_2|\phi\approx 1 $ for moderate volume fraction $\phi = 5\%$: (a) in zero field $\lambda_0\approx 4$, (b) in strong field $\lambda_{\infty}\approx 3$. Hence, when the dipole-dipole interactions are strong enough to compete with the thermal fluctuations, the particles spontaneously aggregate, forming chains of various length (Figure~\ref{fgr:zero_snapshots}). 
On further cooling in zero field the colloid undergoes a hierarchical cascade of structural transformations mediated by sequentially emerging series of topological defects\cite{Kantorovich_Ivanov_Rovigatti_Tavares_Sciortino_2015}, which will result in the percolation of the system of branched clusters (Figure~\ref{fgr:Szero_snapshots}, Supplemental Materials) \cite{Rovigatti_Russo_Sciortino_2012}. 

\begin{figure}[!h]
\centering
    \includegraphics[width=0.495\linewidth]{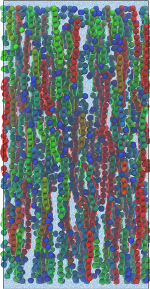}\quad
    \includegraphics[width=0.29\linewidth]{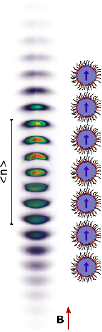}
  \caption{Left: simulation snapshots (box size $24\sigma\times 48\sigma\times 24\sigma$) in strong vertical external field at a volume fraction $\phi=5\%$ and dipole interaction parameter $\lambda=4$ in equilibrium. Particles color-coded by chain length $n$: monomers ($n=1$) are blue, $n\geq10$ - red. Middle: corresponding 3D PDF $P\left(\bm{r}\right)$ in equilibrium. The bar shows the average chain length $\left\langle n\right\rangle\approx3.2$ for comparison. Right: schematic representation of a chain. }
  \label{fgr:inf_snapshots}
\end{figure}

Under a strong applied field the clusters are aligned and unbranched (Figure~\ref{fgr:inf_snapshots}). The corresponding monomer density distribution displays a strongly anisotropic needle-like structure composed of narrow maximums in the monomer positional probability, which are spaced approx. $0.5\sigma$ apart due to the contributions from the chains with odd and even number of particles. The maximum probability is concentrated in the immediate vicinity of the center of mass of the aggregate, which corresponds on average to relatively short chains containing a few particles. The diffuse tail of the PDF persists significantly beyond the few central maxima. 

\begin{figure}[!h]
\centering
  \setlength{\tabcolsep}{0pt}
  \begin{tabular}{ccr}
    \includegraphics[width=0.4875\linewidth]{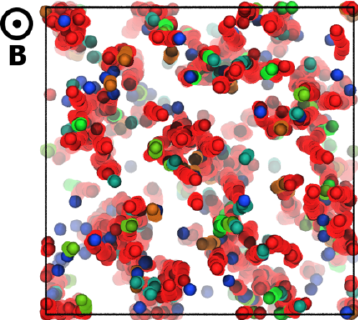}&
    \includegraphics[width=0.4375\linewidth]{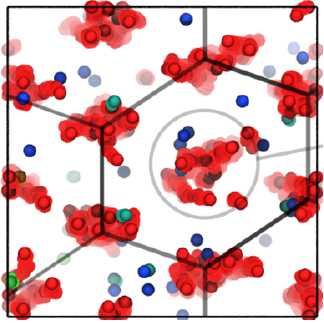}&\enspace\thinspace
    \includegraphics[width=0.05\linewidth]{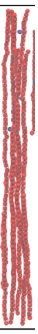}\\
    \textit{(a)}&
    \textit{(b)}&
    \textit{(c)}
  \end{tabular}
  \caption{Simulation snapshots (box size $24\sigma\times 72\sigma\times 24\sigma$) of equilibrium cluster formation of dipolar particles in a strong vertical applied field $\bm{B}$ (viewed from the direction of the field) at a volume fraction $\phi=5\%$, showing the lateral condensation of chains and the emergence of hexagonal order: \textit{(a)} $\lambda=5$, \textit{(b)} $\lambda=7$, \textit{(c)} lateral view of a single strand at $\lambda=7$. Particles color-coded by aggregate size $n$: monomers ($n=1$) are blue, $n\geq10$ - red. }
  \label{fgr:l57_snaphots}
\end{figure}

At small lateral separation the interaction potential $\sim e^{-2\pi r}$ between straight dipolar chains of infinite length decays exponentially with $r$ - the interchain distance \cite{Halsey_Toor_1990,Furst_Gast_2000}. The parallel chains align with an offset by half a diameter $\sigma$, by which they can produce a short-range attraction confined to the near field ($r<2\sigma$) \cite{Halsey_Toor_1990,Halsey_1992}. 
Finite size corrections \cite{Gross_Kiskamp_1997,Furst_Gast_2000} amount to repulsive monopoles bound at the chain ends. The flexibility \cite{Gross_Kiskamp_1997} or an offset by half the chain length \cite{Furst_Gast_2000} and polydispersity of the chain ensemble minimize the repulsion of the chain ends and enable the chain-chain approach into the near field (Figure~\ref{fgr:l57_snaphots}). The far field lateral attraction  $\propto n r^{-5}$ between the two field-aligned $n$-particle chains is originated by the dipole density fluctuations \cite{Halsey_Toor_1990,Furst_Gast_2000} resulting in the condensation of chains into tapered bundled structures (Figure~\ref{fgr:l57_snaphots}c). 
The zero-temperature $\lambda\rightarrow \infty$ ground-state for the ensemble of fully aligned dipolar particles should be a body-centered tetragonal (bct) solid 
\cite{Tao_Sun_1991} comprised by chains offset in the direction of the field  \cite{Chen_Zitter_Tao_1992,Yethiraj_Wouterse_Groh_vanBlaaderen_2004,Zhou_Wen_Sheng_1998}. At ($\lambda>4-7$) the phase diagram of aligned dipolar particles shows a chain and bct phase coexistence \cite{Hynninen_Dijkstra_2005b}. 


\paragraph{Chain length distribution.} 
The chain ($n$-mer) length distribution has been calculated using chemical kinetics \citep{Jordan_1973} or the density functional approach \citep{Zubarev_Iskakova_2000}. According to the later method \citep{Zubarev_Iskakova_1995, Osipov_Teixeira_TelodaGama_1996, Morozov_Shliomis_2002} the chain length distribution $g_n$ should provide the minimum of the free energy density functional of the polydisperse ensemble of chains:  
\begin{equation}
  F = k T \sum_{n=1}^{\infty}g_n \ln\frac{g_n v_p}{e}+g_n f_n \label{eq:F}
\end{equation}
where the first term corresponds to the entropy of an ideal mixture of $n$-mers and the second absorbs intra-chain correlations and interaction with the external field \citep{Zubarev_Iskakova_1995, Osipov_Teixeira_TelodaGama_1996}. The dimensionless internal free energy is $f_n = -\ln Z_n$, where $Z_n$ is the partition function of a self-assembled chain \eqref{eq:SZn} (Supplemental Materials). The minimization of eq.~\eqref{eq:F} is provided by\citep{Zubarev_Iskakova_1995, Osipov_Teixeira_TelodaGama_1996}
\begin{equation}
  g_n = \frac{X^n}{v_p} e^{-\epsilon_*} \label{eq:gn_theor}
\end{equation}
where $\epsilon_* =\ln Z_2$ and $-\epsilon_* k T$ is the average bond energy \citep{Osipov_Teixeira_TelodaGama_1996} accounting for the dispersive thermal fluctuations ($\epsilon_*=2\lambda$ in the zero-temperature limit\citep{Zubarev_Iskakova_2002}) The coefficient $X$ is determined from the condition of particle conservation 
\begin{gather}
  \sum_{n=1}^{\infty}n g_n = \frac{\phi}{v_p} \label{eq:conserv}\\
  X = \frac{1+2\phi \exp{\epsilon_{*}}-\sqrt{1+4\phi\exp{\epsilon_{*}}}}{2\phi\exp{\epsilon_{*}}} \label{eq:X}
\end{gather}
The average number of grains in a chain is \citep{Zubarev_Iskakova_2002} 
\begin{equation}
  \left\langle n \right\rangle = \frac{\sum_{n=1}^{\infty} n g_n}{\sum_{n=1}^{\infty} g_n} = \frac{1}{2}+\sqrt{\frac{1}{4}+\phi Z_2} \label{eq:navg_gn}
\end{equation}
The pair partition function $Z_2$ in zero and infinite field has been calculated \citep{Morozov_Shliomis_2002} \eqref{eq:SZ2zero}-\eqref{eq:SZ2infty} (Supplemental Materials). The asymptotes for $\lambda \gg 1$ \citep{Zubarev_Iskakova_1995, Zubarev_Iskakova_2002} 
\begin{gather}
Z_2\left(\infty\right)\approx\frac{e^{2\lambda}}{3\lambda^2},\quad
Z_2\left(0\right)\approx\frac{1}{\lambda}Z_2\left(\infty\right) \label{eq:Z2_thermal}
\end{gather}
coincide with the second virial coefficients \cite{Gennes_Pincus_1970}. The equations \eqref{eq:Z2_thermal} provide thermal corrections to the zero-temperature behavior $\epsilon_*=2\lambda$ by renormalizing the energy of the bond to account for the thermal fluctuations, which effectively inflate the chain. The predictions of the chain model at low to moderate $\lambda$ significantly depend on the nature of the dependence $\epsilon_{*}\left(\lambda\right)$. To determine the influence of the thermal fluctuations, we exploit the effective parameter $\lambda_*$: $\epsilon_*=2\lambda_*$, where $\lambda_*$ is used to fit the theoretical model to the simulation data and characterizes the effective structuring ability of the dipolar bond.

\begin{figure}[!h]
\centering
  \setlength{\tabcolsep}{0pt}
  \begin{tabular}{cc}
    \raisebox{-0.9\height}{\includegraphics[width=0.8\linewidth]{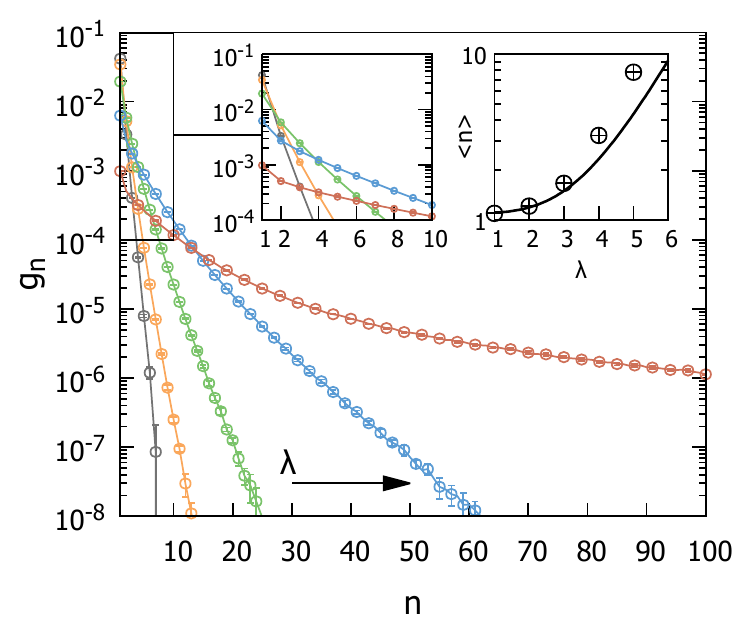}}&\textit{(a)}\\
    \raisebox{-0.9\height}{\includegraphics[width=0.8\linewidth]{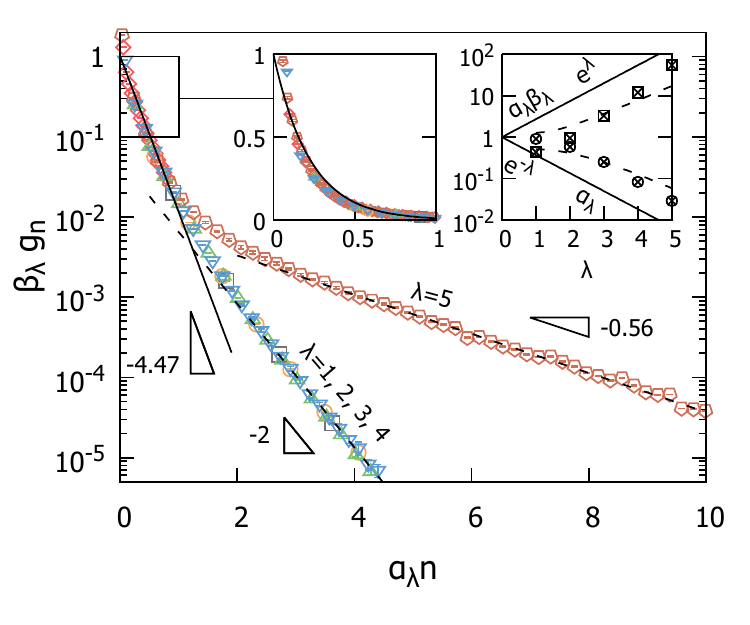}}&\textit{(b)}\\
  \end{tabular}
  \caption{Top: $\left(a\right)$ - equilibrium chain length distributions $g_n$ in strong external field calculated from simulations with dipolar interaction parameter $\lambda=1,2,3,4,5$ ($\phi=5\%$). The left inset shows a detailed view of the initial region. The right inset shows the corresponding average chain length $\left\langle n\right\rangle$: from simulations ($\oplus$) and according to eq.~\eqref{eq:navg_gn} (solid line). Bottom: $\left(b\right)$ - chain length master curve obtained by plotting the equilibrium chain length distribution $g_n$ for various $\lambda$ in the coordinates $\left(\alpha_{\lambda}n,\beta_{\lambda}g_{n}\right)$. Deviations of the tail of $g_n$ from the master curve are observed at larger $\lambda$. Symbols correspond to simulations, dashed lines are intended to guide the eye. The solid line corresponds to the theoretical prediction of the chain model \eqref{eq:gn_theor} and \eqref{eq:alambda_blambda}. The left inset shows a detailed view of the initial region. The right inset shows the dependence of the scaling factors $\alpha_{\lambda}$ and $\beta_{\lambda}$ on $\lambda$ (note: the vertical scaling factor is plotted as $\alpha_{\lambda}\beta_{\lambda}$): obtained by collapsing the simulation data ($\otimes$, $\boxtimes$) and the theoretical predictions of eq.~\eqref{eq:alambda_blambda}; thermal corrections to the factors according to eq.~\eqref{eq:Z2_thermal} are indicated by dashed lines.}
  \label{fgr:gn_eq}
\end{figure}

Under a strong field the particle chains are aligned and their length is increased by approximately a factor of $\propto\sqrt{\lambda}$ \eqref{eq:navg_gn}.
The calculated chain length distribution \eqref{eq:gn_prob} shows that very long chains become accessible at larger $\lambda$ at the cost of the reduced number of free floating monomers and shorter chains. The $n$-mer length distribution smoothly persists across many orders of magnitude and has an exponential character as predicted by eq.~\eqref{eq:gn_theor}. There is a transition between two exponentially distributed regimes with different exponents, which is more apparent for larger $\lambda$. The deviation from the single exponential decay for longer chains has been observed experimentally \citep{Klokkenburg_Dullens_Kegel_Erne_Philipse_2006}. 
The average length of a chain calculated using $g_n$ \eqref{eq:navg_gn} shows good agreement with the predictions of eqs.~\eqref{eq:gn_theor}-\eqref{eq:Z2_thermal} (Figure~\ref{fgr:gn_eq}a, right inset). 

The appropriate scaling for $g_n$ follows from equation \eqref{eq:gn_theor} and a master curve can be obtained by plotting in rescaled coordinates $\left(\alpha_{\lambda}n,\beta_{\lambda}g_{n}\right)$ (Fig.~\ref{fgr:gn_eq}b), where 
\begin{equation}
\alpha_{\lambda}=-\sqrt{\phi}\ln{X},\quad\beta_{\lambda}=e^{\epsilon_{*}}
\label{eq:alambda_blambda}
\end{equation}
For $\lambda>4$ only the first exponential regime collapses satisfactorily, whereas the tail of the $g_{n}$ progressively deviates, reflecting the lateral condensation of the chains, forming thick bundles (Fig.~\ref{fgr:l57_snaphots}). For $\epsilon_{*}\gg 1$ to leading order $\alpha_{\lambda}\approx e^{-0.5\epsilon_{*}}$. The scaling coefficients estimated from collapsing the simulation data are plotted in the right inset of Fig.~\ref{fgr:gn_eq}b. The asymptotes are provided by the zero temperature limit $\epsilon_{*}=2\lambda$, the thermal corrections \eqref{eq:Z2_thermal} are indicated by the dashed lines. The scaling coefficients gradually approach the predicted limiting behavior as $\lambda$ is increased, but the influence of the thermal fluctuations is substantial in the whole investigated range.

\subsubsection{Non-equilibrium}
\label{sec:non-equilibrium}
Simple shear $\bm{v}=\left(\dot{\gamma} y,0,0\right)$ is the main rheometric flow governed by a single viscosity coefficient. The majority of the rheological data is reported in terms of its shear rate $\dot{\gamma}$ dependence. 
Figure~\ref{fgr:Sinf_snapshots} (Supplemental Materials) presents a simulation snapshot of the field structured colloid in simple shear and the corresponding PDF is shown in a projection onto the shearing plane $P\left(x,y\right)=\int{P\left(\bm{r}\right)dz}$ (Figure~\ref{fgr:Spdf_chain}). Qualitatively, the hydrodynamic stress erodes the microstructure: the number of longer chains is reduced in comparison to the equilibrium state, which is also accompanied by the visible collapse of the tail of the PDF under shear and a slight decrease of the average chain length. The deviation of the chains from their alignment in the direction of the field is reflected in the deformation of the PDF.

\begin{figure}[!h]
\centering
  \includegraphics[width=0.7\linewidth]{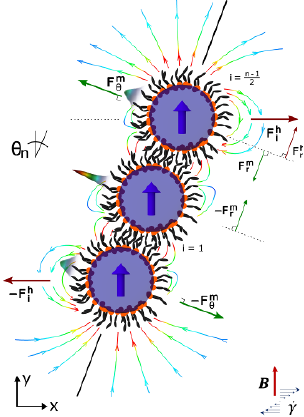}
  \caption{Schematic representation of a straight chain of dipolar particles (here a trimer containing $n=3$ particles) with field-aligned dipoles and acting forces in simple shear and strong magnetic field applied vertically, magnetic field lines around the chain (note: vector lengths not to scale). The particles are indexed from the center ($i=1$). The peaks of the positional probability distribution $P_3$ along the director show that the central particle is well localized near the center of mass, whereas the end particles are less confined. }
  \label{fgr:theor_scheme}
\end{figure}

The microstructural model of chain-forming dipolar colloids is constructed based on the balance of the micromechanical stress\cite{Zubarev_Iskakova_1995,Yu_Zubarev_Yu_Iskakova_2007}.
In the chain model\cite{Yu_Zubarev_Yu_Iskakova_2007} the aggregates of dipolar particles are approximated by a straight rod-like chain (Figure~\ref{fgr:theor_scheme}), comprising particles in close contact, spaced equidistantly along the chain and the axis (director $\bm{e}_r$) of the linear $n$-mer forms an angle $\theta_n$ with the direction of the field. Here we assume a complete alignment of the individual dipole moments in the direction of the applied field.
The total torque balance determines the orientational response of the rigid $n$-particle rod to the action of the hydrodynamic shear ($\bm{\Gamma^h_n}$) and polarization ($\bm{\Gamma^m_n}$)
\begin{equation}
  \bm{\Gamma_n^m}+\bm{\Gamma_n^h}=0
\end{equation}
Tallying the dipolar ($\bm{F^m}$) and hydrodynamic ($\bm{F^h}$) force couples starting from the center ($i=1$) of the chain\citep{deVicente_Lopez-Lopez_Duran_Gonzalez-Caballero_2004}: $\bm{\Gamma^{m,h}}=2\sum_{i=1}^{\nu_n} i\sigma\bm{e_{r}} \times\bm{F_{i}^{m,h}}$ where  $\bm{e_{r}}$ is the director vector of the chain and $\nu_n=\frac{n-1}{2}$.

The energy of the magnetic bond $U_{dd}\left(\bm{e_{y}},\bm{e_{y}},\bm{e_{r}}\right)=-\lambda k T \left( 3 \cos^{2}\theta_n - 1 \right)$ produces a magnetic force couple: $\bm{F_{\theta}^{m}} = -\frac{1}{\sigma}\frac{\partial U_{dd}}{\partial \theta}\bm{e_{\theta}} = -6\lambda k T\sigma^{-1} \cos\theta_n\sin\theta_n \bm{e_{\theta}}$ and $-\bm{F_{\theta}^{m}}$ acting on the i\textsuperscript{th} and (i-1)\textsuperscript{th} particles. Summing the individual contributions along the chain, a realigning torque is provided by the end particles, where the magnetic force couples are not compensated $\bm{\Gamma^{m}}= -6\left(n-1\right)\lambda k T \cos\theta_n\sin\theta_n \bm{e_{r}}\times\bm{e_{\theta}}$.

In turn, assuming the free draining limit, the hydrodynamic force acting on the i\textsuperscript{th} particle of the $n$-particle chain in simple shear $\bm{F^{h}_{i}}=\beta i \dot{\gamma} \sigma \cos\theta_n \bm{e_{x}}$ and the deviating torque $\bm{\Gamma^{h}}=\frac{1}{12}n \left(n^2-1\right) \beta \dot{\gamma} \sigma^2 \cos^2\theta_n \bm{e_{r}} \times\bm{e_{\theta}}$.

The torque equilibrium yields the deviation angle of the $n$-mer
\begin{equation}
  \tan\theta_n=\frac{1}{12}\left(n^2+n\right)Mn \label{eq:tantheta_Mn}
\end{equation}

The typical aggregation/disintegration scenario is tentatively governed by eq.~\eqref{eq:tantheta_Mn}, whereby the chains gradually grow in length by attaching monomers and other free-floating segments and deviate quadratically with $n$ from the direction of the field. The fragmentation is induced by the failure of the bond vicinal to the center of the chain, where the highest tensile stress is applied \cite{Martin_Anderson_1996}, and the fragments realign.

\paragraph{Chain erosion.} 

In shear flow the longer chains are destroyed by the hydrodynamically applied tensile stress. The simulations show that increasing the shear rate $\dot{\gamma}$ the chain length distribution is gradually eroded at $n\gg 1$, however, retaining the double-exponential form. This behavior suggests that a similar scaling may pertain and a universal curve can be obtained in a similar way as previously by collapsing the non-equilibrium chain-length distribution $g_n$ onto the equilibrium distributions at vanishing shear rates $Mn\rightarrow0$ for each $\lambda$, which are then collapsed onto a single master curve for all $\lambda$ using the previously determined equilibrium scaling factors $\alpha_{\lambda}$ and $\beta_{\lambda}$ \eqref{eq:alambda_blambda} (Figure~\ref{fgr:gn_eq}). 
\begin{figure}[]
\centering
  \includegraphics[width=0.8\linewidth]{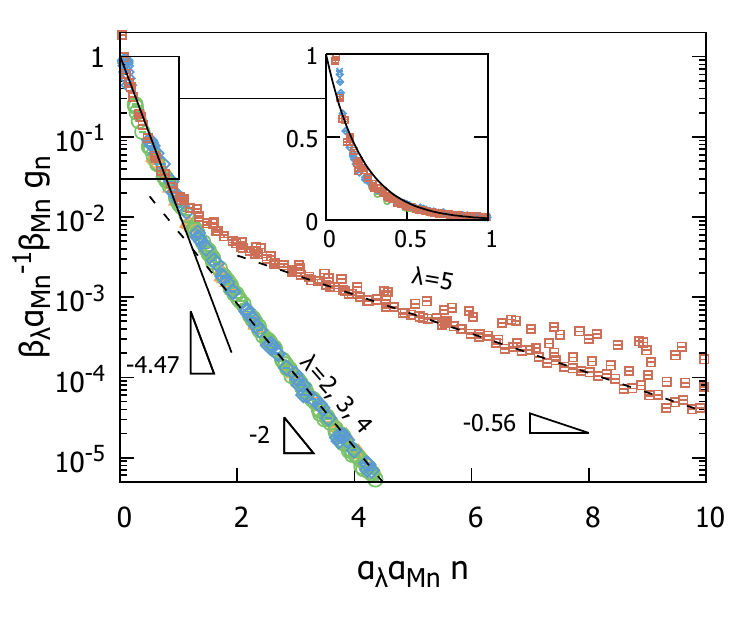}
  \caption{  
 Non-equilibrium chain length distribution $g_n$ in strong applied field for $\phi=5\%$ and different interaction strengths and shear rates: $\lambda=2$ ($Mn=0-1.33$), $\lambda=3$ ($Mn=0-0.88$), $\lambda=4$ ($Mn=0-0.028$), $\lambda=5$ ($Mn=0-0.52$) plotted in the scaled coordinates $\left(\alpha_{\lambda}\alpha_{Mn}n, \beta_{\lambda}\alpha_{Mn}^{-1}\beta_{Mn}g_n\right)$. Using the scaling factors $\alpha_{Mn}$ and $\beta_{Mn}$ for each $\lambda$ the non-equilibrium distributions are collapsed onto the equilibrium $g_n$ at $Mn=0$ (compare with Figure~\ref{fgr:gn_eq}b). Inset shows the detailed view of the initial region. The solid line is the same as in Figure~\ref{fgr:gn_eq} and represents the prediction of the chain model. }
  \label{fgr:MCMn}
\end{figure}
In the result, Figure~\ref{fgr:MCMn} shows the chain-length distributions $g_n$ calculated from the simulation data for various $\lambda=2-5$ and $Mn$ and plotted in the coordinates $\left(\alpha_{\lambda}\alpha_{Mn}n, \beta_{\lambda}\alpha_{Mn}^{-1}\beta_{Mn}g_n\right)$, where $\alpha_{Mn}$ and $\beta_{Mn}$ are the corresponding non-equilibrium horizontal and vertical scaling factors (Note: the vertical scaling is achieved by the factor $\alpha_{Mn}^{-1}\beta_{Mn}$). 
At $\lambda=4$ we observe an anomalous behavior for some shear rates from $Mn\approx 0.03$ to $Mn\approx 0.15$ (Fig.~\ref{fgr:Sg4_Mn}, Supplemental Materials), where instead of being eroded, the tail of the distribution $g_{n}$ recovers with increasing shear rate. This evidences that a magnetohydrodynamic instability is in effect and the system enters a lamellar phase (see also Section~\ref{sec:phase_diagram}). In this regime the chains of nanoparticles rearrange themselves to form mesoscopic layers in the field-velocity plane. The associated shear-banding reduces the hydrodynamic stresses within the layered phase and the chains gradually extend. The Mason number where this happens is on the very edge of the experimental window for the magnetic colloids. In turn, for higher interaction strength $\lambda=5$ the phase boundary shifts towards lower shear rates and the delamination occurs before the shear flow can significantly erode the chain length distribution, hence, only a slight dispersion of the collapsed data is observed in Figure~\ref{fgr:MCMn}. Therefore, the same scaling as in equilibrium is also useful in the lamellar phase. 

\begin{figure}[h]
\centering
  \includegraphics[width=0.495\linewidth]{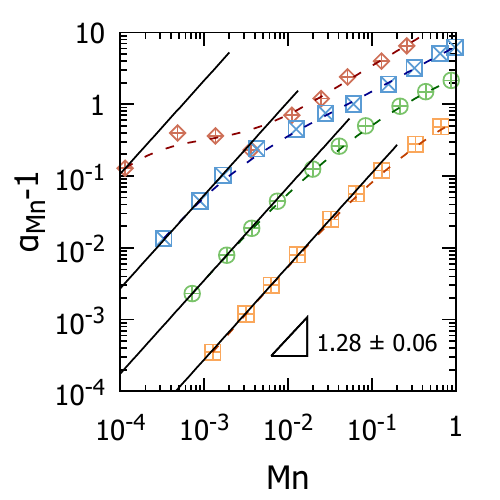}
  \includegraphics[width=0.495\linewidth]{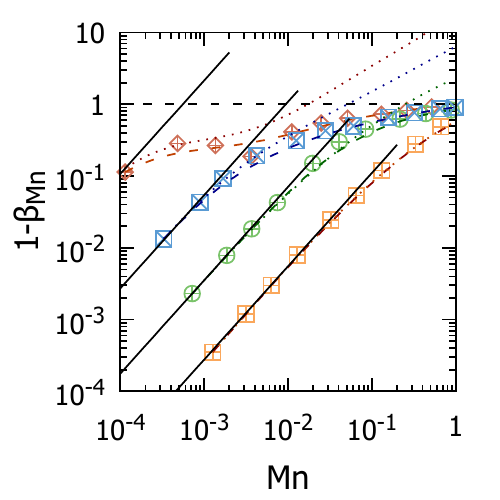}
  \caption{Scaling factors $\alpha_{Mn}$ and $\beta_{Mn}$ used to collapse the non-equilibrium simulation data in Figure~\ref{fgr:MCMn} as a function of the Mason number $Mn$ for $\lambda=2$ ($\boxplus$), $\lambda=3$ ($\oplus$), $\lambda=4$ ($\boxtimes$), $\lambda=5$ ($\diamondplus$). The lines are intended to guide the eye.}
  \label{fgr:MCMn_abMn}
\end{figure}

The scaling factors $\alpha_{Mn}$ and $\beta_{Mn}$ are plotted in Figure~\ref{fgr:MCMn_abMn} as a function of the Mason number $Mn$ for various investigated interaction strengths $\lambda=2-5$. We noticed that $\alpha_{Mn}\cdot\beta_{Mn}\approx 1$ and $\alpha_{Mn}-1\approx1-\beta_{Mn}$ for lower shear rates $Mn<10^{-2}$. The chain model, however, does not support the construction of the chain length master curve in non-equilibrium, so it is difficult to interpret this behavior.

Considering the radial force balance of a rigid chain (Figure~\ref{fgr:theor_scheme}): the $n$-particle chain is broken, when the strength of the bond $F_{r}^{m} = -\frac{\partial U_{dd}}{\partial r}=-3\lambda kT \sigma^{-1}\left(3\cos^2\theta_n-1\right)$ is overcome by the overall shear force extending the chain $F^{h}_{r}=\sum_{i=1}^{\nu_n} \bm{F_{i}^{h}\cdot\bm{e_{r}}}=\frac{1}{8}\left(n^2-1\right) \beta \dot{\gamma} \sigma \cos\theta_n \sin\theta_n$. The equation for the maximum length of a stable chain is produced from the radial mechanical balance of the bonding force versus the hydrodynamic erosive force $F_{r}^{m}+F_{r}^{h}=0$ accounting for eq.~\eqref{eq:tantheta_Mn}: $4n^4+5n^3-2n^2-3n-\frac{288}{Mn^2}=0$. To a leading order in $n$ the critical length is
\begin{equation}
n_{crit}\approx\sqrt{\frac{6\sqrt{2}}{Mn}} \label{eq:ncrit}
\end{equation}
Hence, any $n$-mer with $n>n_{crit}$ is broken up by the shear flow. 

\begin{figure}[!h]
\centering
  \includegraphics[width=0.495\linewidth]{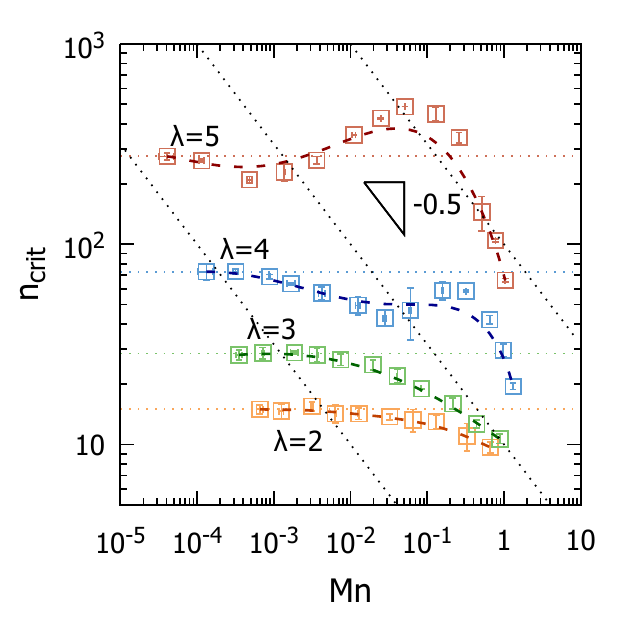}
    \includegraphics[width=0.495\linewidth]{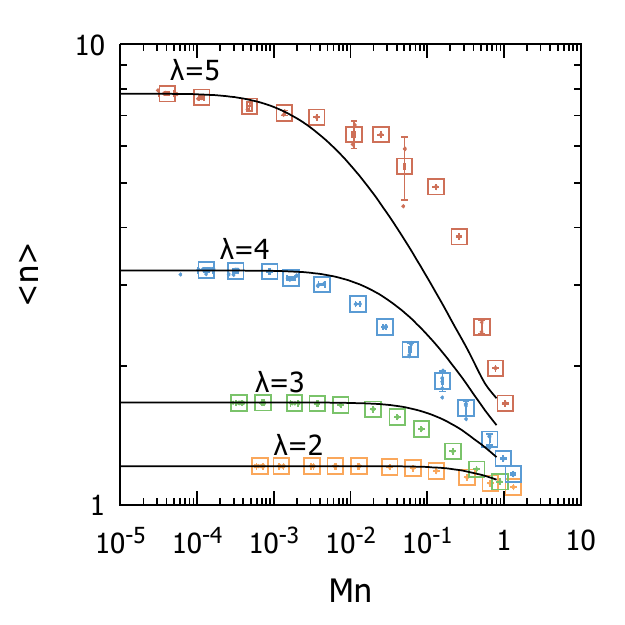}
  \caption{Left: maximum number of particles in a single aggregate observed in simulations at $\lambda=2,3,4,5$ ($\phi=5\%$). Non-monotonic behavior at larger $\lambda$ evidences the occurring microhydrodynamic instabilities. Dashed lines are intended to guide the eye. Right: average number of particles in the chain $\left\langle n\right\rangle$ calculated from the non-equilibrium chain length distribution $g_n$ according to eq.~\eqref{eq:navg_gn} and plotted as a function of the Mason number $Mn$ for $\lambda=2,3,4,5$ ($\phi=5\%$): symbols are simulation data ($\Box$), solid lines are theoretical predictions of the chain model for effective dipole interaction parameter $\lambda^{\left\langle n\right\rangle}_{*}=0.82,1.55,2.483,3.484.$}
  \label{fgr:ncrit_navg}
\end{figure}

Figure~\ref{fgr:ncrit_navg} shows the maximum number of particles in an identified aggregate that has been observed in a simulation trajectory at various interaction strengths $\lambda$ and shear rates $Mn$. This quantification is not exact, because it will depend on the duration of the production run, but considering the exponential decay of the chain length distribution it provides a good estimate of the actual critical length. The proportionality to $Mn^{-0.5}$ of the critical length $n_{crit}$ predicted by eq.~\eqref{eq:ncrit} is not observed. In fact, we observe a non-monotonous dependence of $n_{crit}$ on the shear rate $Mn$ at some $\lambda\geq 4$, which is consistent with the transition into the shear-banded state and the associated decrease of the erosive hydrodynamic stresses within the layered phase. At $\lambda<4$ we indeed observe a power law decay $n_{crit}\propto Mn^{-p}$ but with a lower exponent than the predicted $0.5$ \eqref{eq:ncrit}.

The concept of a critical chain \cite{Martin_Anderson_1996, Zubarev_Fleischer_Odenbach_2005} enables to calculate the non-equilibrium chain length distribution in an empirical way\cite{Yu_Zubarev_Yu_Iskakova_2007}. The distribution $g_{n}$ can be estimated in simple shear by carrying out minimization \eqref{eq:gn_theor} numerically with $\epsilon_{*}=-\beta U_{dd}\left(\theta_{n}\right)=\lambda\left(3\cos^2{\theta_n}-1\right)$ under the conservation condition \eqref{eq:conserv}, introducing the critical length \eqref{eq:ncrit} as an upper bound in eq.~\eqref{eq:conserv}. The use of the global extremal principle is justified in the vicinity of the equilibrium, which for magnetic colloids corresponds to the whole range of the practically accessible shear rates $Mn\ll1$.

Figure~\ref{fgr:ncrit_navg} shows the average number of particles in a chain \eqref{eq:navg_gn} under shear and in strong applied field. For all investigated $\lambda=2-5$ the plateau, corresponding to the low shear rate regime $Mn\rightarrow 0$, is followed at some critical $Mn>10^{-2}$ by a near-power law decrease of the average chain length $\left\langle n\right\rangle$, which evidences a rapid breakdown of the microstructure. The theoretical predictions of the non-equilibrium chain model were obtained by fitting the theory to the simulation data in the plateau region, corresponding to the average chain length in equiilibrium, using a fitting parameter $\lambda_*^{\left\langle n\right\rangle}$, i.e. $\epsilon_*=2\lambda_*^{\left\langle n\right\rangle}$. 
The result is shown in comparison to the simulation data in Figure~\ref{fgr:ncrit_navg} exhibiting a good agreement.
The theoretical model is not strictly applicable for $\lambda\geq 5$ because of the observed side to side aggregation of the chains forming thick bundles. Still, even in this regime it fairly well reproduces the behavior observed in the simulations and can be used with the appropriate selection of the effective interaction parameter $\lambda_*^{\left\langle n\right\rangle}$.

\paragraph{Chain deviation.} The field- and shear-induced conformational anisotropy of the microstructure bonded by the dipolar interactions is contained in the gyration tensor of the $n$-mer
$G_{ij}^n=\frac{1}{n}\sum_{k=1}^n\left\langle r^{\left(k\right)}_i r^{\left(k\right)}_j \right\rangle$, where $r_{i}^{\left(k\right)}$ is the i\textsuperscript{th} component of the position vector $\bm{r^{\left(k\right)}}$ of the k\textsuperscript{th} particle relative to the center of mass of the $n$-mer. It is related to the positional probability distribution function $P_n\left(\bm{r}\right)$ through
\begin{gather}
G_{ij}^n=\frac{\phi}{nv_pg_n}\int_V r_{i}r_{j}P_n\left(\bm{r}\right)dV
\end{gather}
The deviation of the chain from the vertical ($y$) direction $\theta_n$ is obtainable from the gyration tensor
\begin{gather}
\tan\left(2\theta_n\right)=\frac{2G_{xy}^n}{G_{yy}^n-G_{xx}^n} \label{eq:tangyr}
\end{gather}
In zero field a polymer-like behavior \cite{Aust_Kroger_Hess_1999} is observed: under shear the spherical symmetry of the PDF is reduced (Figure~\ref{fgr:Szero_prob}, Supplemental Materials), the monomer probability distribution is compressed in the velocity gradient ($y$) direction and extended in the flow ($x$) direction. Thereby the averaged conformations resemble ellipsoids and the average steady alignment is induced by the flow. The conformations (birefringence) ellipsoid is oriented at an average angle $\left\langle \theta\right\rangle $ (measured relative to the direction of the velocity gradient). 
We observe signs of the limiting behavior $\left\langle \theta\right\rangle\rightarrow 45^{\circ}$ (extinction angle) for vanishing shear rates $Mn\rightarrow 0$\citep{Philippoff_1956}. For finite shear rates the chains are gradually aligned with the flow ($\left\langle \theta\right\rangle\rightarrow 90^{\circ}$).

\begin{figure}[!h]
\centering
    \includegraphics[width=0.7\linewidth]{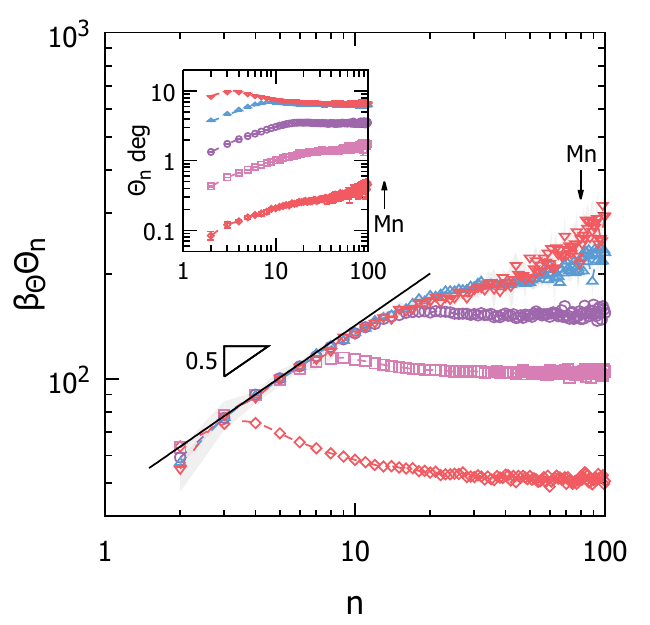}
  \caption{Scaled chain deviation angle $\beta_{\theta}\theta_n$ from the direction of the applied field as a function of the chain length $n$ for $\lambda=5$, volume fraction $\phi=5\%$ and different Mason numbers $Mn=4.1\cdot 10^{-5}$ ($\bigtriangledown$), $4.8\cdot 10^{-4}$ ($\bigtriangleup$), $3.6\cdot 10^{-3}$ ($\circ$), $0.025$ ($\Box$), $0.13$ ($\diamond$). Solid line is a power law $\theta_n\propto\sqrt{n}$. The inset shows the original unscaled data $\theta_n$. }
  \label{fgr:MCtheta}
\end{figure}

In the case of a strong vertically applied field, according to eq.~\eqref{eq:tantheta_Mn}, the chain-length distribution $g_{n}$ produces the distribution of the deviation angle $\theta_{n}$ due to a significant quantity of subcritical chains with $n<n_{crit}$. The angle distributions obtained from the simulation data at different shear rates (shown in Figure~\ref{fgr:MCtheta} for $\lambda=5$ and is similar for all calculated $\lambda$) can be partially collapsed onto a single master curve when plotted in the coordinates $\left(n,\beta_{\theta}\theta_n\right)$, where $\beta_{\theta} $ is a coefficient, which depends on $Mn$, as predicted by eq.~\eqref{eq:tantheta_Mn}.

\begin{figure}[!h]
\centering
    \includegraphics[width=0.7\linewidth]{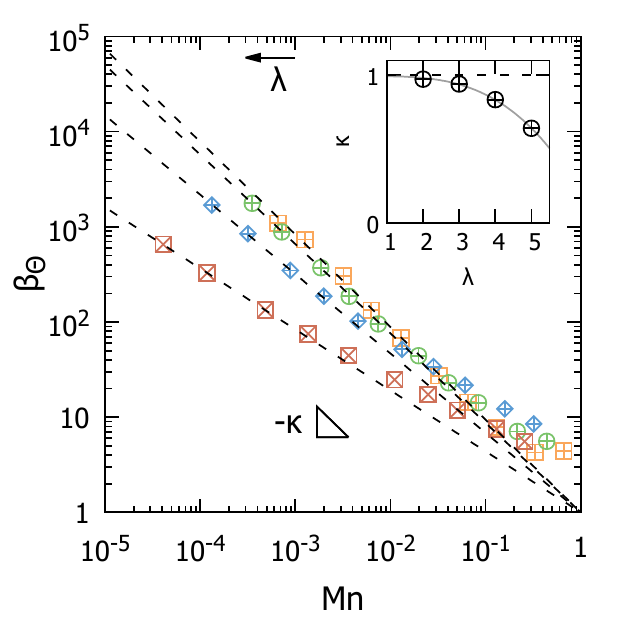}
  \caption{Dependence of the scaling factor $\beta_{\theta}$ in Fig.~\ref{fgr:MCtheta} on the Mason number $Mn$ for various dipole interaction parameters $\lambda=2,3,4,5$. The dashed lines are power law fits $\propto Mn^{-\kappa}$ to the simulation data for $\lambda=2$ ($\boxplus$), $\lambda=3$ ($\oplus$), $\lambda=4$ ($\diamondplus$), $\lambda=5$ ($\boxtimes$). Inset: dependence of the exponent $\kappa$ on $\lambda$. The limiting behavior $\kappa\rightarrow 1$ is consistent with eq.~\eqref{eq:tantheta_Mn}. The lines are intended to guide the eye. }
  \label{fgr:MCtheta_bMn}
\end{figure}

\begin{figure}[!h]
\centering
  \includegraphics[width=0.495\linewidth]{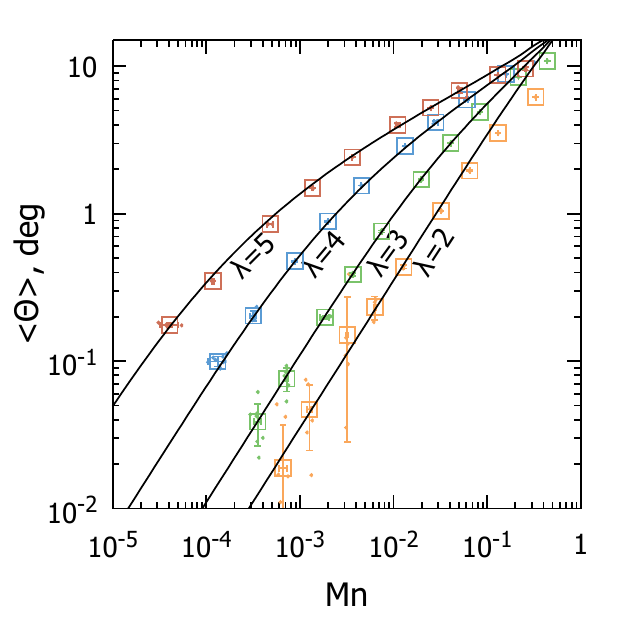}
  \includegraphics[width=0.495\linewidth]{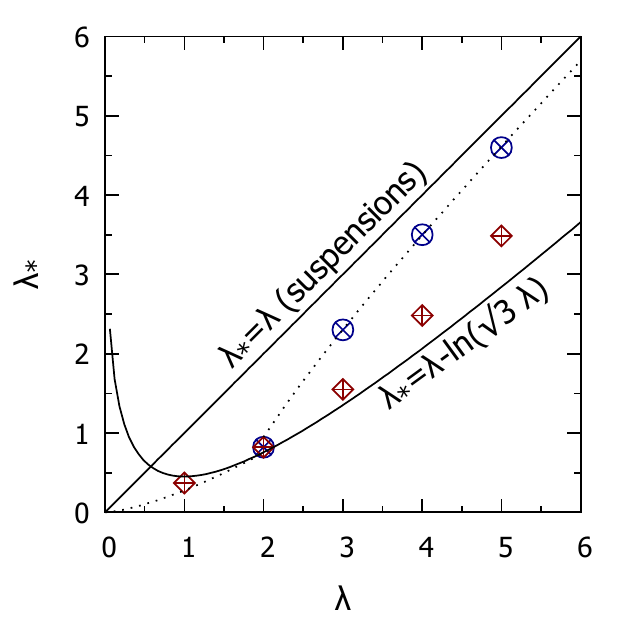}
  \caption{Left: average deviation angle $\left\langle\theta\right\rangle$ from the direction of the applied field calculated according to eq.~\eqref{eq:theta_avg} as a function of the Mason number $Mn$ for various $\lambda$ ($\phi=5\%$): symbols represent simulation data for $\lambda=2,3,4,5$; lines - predictions of the chain model for effective dipole interaction parameter $\lambda^{\left\langle\theta\right\rangle}_{*}=0.82,2.3,3.5,4.6$. Right: effective dipole interaction parameter $\lambda_{*}$ used to adjust the predictions of the chain model to the simulation data as a function of the nominal $\lambda$: $\lambda^{\left\langle n\right\rangle}_{*}=0.37,0.82,1.55,2.483,3.484.$ obtained by fitting to the average chain length $\left\langle n\right\rangle$ ($\diamondplus$) and to the average deviation angle $\left\langle\theta\right\rangle$ - $\lambda^{\left\langle\theta\right\rangle}_{*}=0.82,2.3,3.5,4.6$ ($\otimes$). Solid lines show the zero-temperature limit $\lambda_{*}=\lambda$ and thermal corrections of eq.~\eqref{eq:Z2_thermal}. Dotted lines are intended to guide the eye.}
  \label{fgr:theta_avg_neps}
\end{figure}

Initially, $\theta_{n}$ scales as $\propto\sqrt{n}$ in contrast to eq.~\eqref{eq:tantheta_Mn}, where a larger exponent was expected. For greater $n$ gradual deviations from the power law are observed, leading eventually to a plateau where the deviation angle is independent of the chain length but depends on the Mason number $Mn$.
The coefficients $\beta_{\theta}$ used to obtain the partial collapse are plotted in Figure~\ref{fgr:MCtheta_bMn}. The expected near-linear scaling of the deviation angle with respect to the Mason number $\theta_n\propto Mn$ predicted by eq.~\eqref{eq:tantheta_Mn} is observed only for the smallest $\lambda$, whereas for larger $\lambda$ we observe a power law $\theta_{n}\propto Mn^{\kappa}$ with the exponent $\kappa<1$ (plotted in the inset of Figure~\ref{fgr:MCtheta_bMn}), which decreases with increasing $\lambda$. 
The predicted maximum deviation angle of the critical chain is $\theta_{max}\approx35^{\circ}$ (from \eqref{eq:tantheta_Mn} and \eqref{eq:ncrit}). This is not the case in our simulations. The largest angle we observe is $\approx 10^{\circ}$ for the largest shear rates. The observed discrepancies are most likely due to the flexibility of the chains, which is not accounted for in the chain model.
Using the chain length distribution $g_n$ the statistical averaging of the inclination angle can be produced similar to eq.~\eqref{eq:navg_gn}
\begin{equation}
  \left\langle \theta \right\rangle = \frac{\sum_{n=2}^{\infty} \theta_{n} g_n}{\sum_{n=2}^{\infty} g_n} \label{eq:theta_avg}
\end{equation}
The average angle $\left\langle \theta \right\rangle$ is presented in Figure~\ref{fgr:theta_avg_neps} as a function of the Mason number $Mn$. At low shear rates ($Mn\rightarrow 0$) there is a linear regime, where the alignment angle increases proportionally to Mn: $\left\langle \theta \right\rangle\propto Mn$.

In contrast to the usual polymer theory where the angular response is solely due to the orientational influence of the shear flow, the orientational resistance\cite{Aust_Kroger_Hess_1999} has no meaning for the field-structured dipolar colloids, where the angle is determined by the simultaneous competition of the shear and external field. Instead, we define the angular compliance $s$:
\begin{equation}
  \left\langle \theta \right\rangle = s Mn \label{eq:sMn}
\end{equation}

\begin{figure}[]
\centering
  \includegraphics[width=0.8\linewidth]{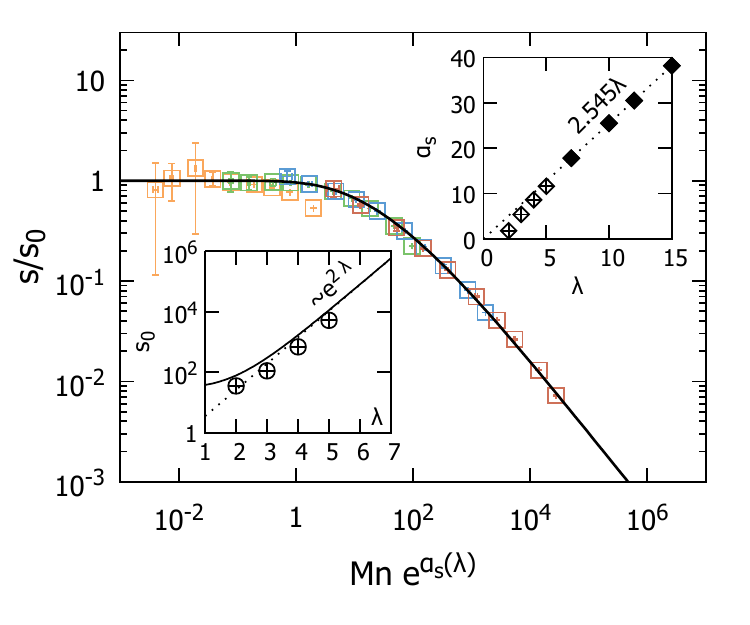}
  \caption{Orientational compliance $s$ defined by eq.~\eqref{eq:sMn} as a function of the Mason number plotted in the scaled coordinates $\left(e^{\alpha_s}Mn,s_0^{-1} s\right)$ ($\phi=5\%$): simulation data for $\lambda=2,3,4,5$ ($\Box$), solid line - master curve predicted by the chain model for $\lambda\gg 1$. Left inset: orientational compliance $s_0$ at vanishing shear $Mn\rightarrow 0$ as a function of $\lambda$ obtained by collapsing the available simulation data ($\oplus$) and the predictions of the chain model (solid line), dotted line is eq.~\eqref{eq:s0}. Right inset: horizontal scaling factor $\alpha_s$ as a function of $\lambda$ obtained by collapsing the available simulation data ($\diamondplus$) and theoretical curves predicted by the chain model for $\lambda\geq 7$ ($\filleddiamond$), dotted line is a fit $\alpha_s=2.545\lambda$.}
  \label{fgr:sMn}
\end{figure}
Figure~\ref{fgr:sMn} shows the orientational compliance calculated according to eq.~\eqref{eq:sMn} and plotted in the coordinates $\left(e^{\alpha_s}Mn,s_0^{-1}s\right)$. For vanishing shear rates and large dipolar strength $\lambda\gg 1$ it can be shown without difficulty that 
\begin{equation}
 s_0 =\left.\frac{\partial \left\langle \theta \right\rangle}{\partial Mn}\right|_{Mn \rightarrow 0} \approx \frac{1}{6}\phi e^{2\lambda}\frac{180^\circ}{\pi}
 \label{eq:s0}
\end{equation}
In practice, this relation holds starting from $\lambda > 4 $ as shown in the lower inset of Fig.~\ref{fgr:sMn}. The simulations show that the horizontal shift factor $\alpha_s$ is approximately proportional to $\lambda$: $\alpha_s\approx 2.545\lambda$. Using these scaling factors a single master curve can be produced for the orientational compliance for all investigated dipolar strengths $\lambda=2-5$ and shear rates $Mn$.

Figure~\ref{fgr:theta_avg_neps} shows the effective interaction strengths $\lambda^{\left\langle n\right\rangle}_{*}$ and $\lambda^{\left\langle\theta\right\rangle}_{*}$, which were used as fitting parameters to connect the predictions of the chain model with the structural parameters measured from the simulations, i.e. the average chain length $\left\langle n\right\rangle$ and the average deviation angle $\left\langle\theta\right\rangle$ calculated at various nominal $\lambda$. A limiting behavior $\lambda_*=\lambda$ corresponds to the rigid chain model\cite{Zubarev_Iskakova_1995,Zubarev_Iskakova_2000} disregarding any influence of the thermal fluctuations, which is indicated by a solid line in Figure~\ref{fgr:theta_avg_neps}. The deviations from the linear dependence can be explained as the influence of the Brownian motion. The second line shows the thermal corrections \eqref{eq:Z2_thermal} for a strong applied field. At $\lambda<2$ the asymptotic behavior \eqref{eq:Z2_thermal} diverges and should not be taken into account. 
The simulations indicate that increasing the nominal interaction parameter $\lambda$ the influence of the Brownian motion on the inclination angle reduces faster than the influence upon the length of the chain, for which $\lambda^{\left\langle n\right\rangle}_{*}$ more or less follows the predicted curve. In fact, the rotational diffusivity $D_{rot}$ of a linear $n$-mer diminishes with larger $n$, i.e. $D_{rot}\propto n^{-1}$. The increase of the interaction strength $\lambda$ leads to the increase of the average chain length. Hence, the particles in the chain will still fluctuate at their positions, whereas the chains themselves are non-Brownian units for $\lambda>5$. Overall, the values of the effective interaction parameters, which were used to calibrate the chain model to the simulation data, are completely within the expected range, i.e. bounded by the zero-temperature limit from above and by the thermal corrections \eqref{eq:Z2_thermal} from below.
To add, the chain-chain interactions, which are unaccounted for in the current model, may also be acting and at least partially responsible for the lower values of the effective parameter $\lambda^{\left\langle n\right\rangle}_{*}<\lambda^{\left\langle \theta\right\rangle}_{*}$. The estimates show that the interchain interactions lead to a decease of the average length of the chains\cite{Iskakova_Zubarev_2002}.

\subsection{Rheological properties}
\label{sec:rheological_properties}
\subsubsection{Stress tensor}
\label{sec:stress_tensor}

The deviatoric stress tensor of the field-structured dipolar colloids comprises the usual symmetric and an additional antisymmetric component, which appears due to the breaking of the rotational symmetry by the imposed field and the associated violation of the angular momentum conservation\cite{Zubarev_Iskakova_1995}
\begin{equation}
\sigma_{ij}=2\eta_{0}\gamma_{ij}+\sigma^{s}_{ij}+\sigma^{a}_{ij} \label{eq:sigma_ij}
\end{equation}
The antisymmetric stress increment $\sigma_a$ is generated by the system of couples induced by the noncentral dipolar interactions and is proportional to the number density $g_n$ of the torque-producing units in the polydisperse ensemble of aligned chains\cite{Martin_Anderson_1996}\cite{Yu_Zubarev_Yu_Iskakova_2007}
\begin{equation}
\sigma^{a}_{xy}=\frac{1}{2}\sum^{n_{crit}}_{n=2}\left(\bm{\Gamma^{m}_{n}}\cdot\bm{e_{z}}\right)g_{n}=\eta_{0}\dot{\gamma}\sum^{n_{crit}}_{n=2}\frac{9}{Mn}\frac{\left(n-1\right)\tan\theta_{n}}{1+\tan^2\theta_{n}}v_pg_{n} \label{eq:sigma_a}
\end{equation}
From eq.~\eqref{eq:tantheta_Mn} $\tan\theta_n\approx\theta_n\propto s_n Mn $, hence, the antisymmetric stress increases linearly with the shear rate generating the low shear pseudo-Newtonian plateau.
In turn, the symmetric stress increment $\sigma_{xy}^{s}$ is of hydrodynamic origin and is similar to a suspension stress, which originates from the perturbation of the shear flow introduced by the presence of the self-assembled microstructure. The explicit calculation of the flow field is only possible for simpler anisometric shapes\cite{Pokrovskii_1972, Brenner_1974} embedded in a fluid, e.g. a rigid ellipsoid circumscribing the aggregate\cite{Shulman_Kordonsky_Zaltsgendler_Prokhorov_Khusid_Demchuk_1986}. Zubarev and Iskakova\cite{Zubarev_Iskakova_1995,Yu_Zubarev_Yu_Iskakova_2007} approximate a linear $n$-particle chain by an ellipsoid of revolution with axes $\left(\sigma, n\sigma\right)$ and provide expressions for the symmetric stress increment of the ensemble of chains with a length distribution $g_n$:
\begin{align}
\sigma_{xy}^{s}=\eta_{0}\dot{\gamma}&\sum_{n=1}^{n_{crit}}nv_pg_{n}\Big[\alpha_{n}+\frac{1}{2}\left(\xi_{n}+\beta_{n}\lambda_{n}\right)\left(\left\langle e_{x}^{2}\right\rangle_{n}+\left\langle e_{y}^{2}\right\rangle_{n}\right) \nonumber \\
& +\frac{1}{2}\beta_{n}\left(\left\langle e_{y}^{2}\right\rangle_{n}-\left\langle e_{x}^{2}\right\rangle_{n}\right)+\left(\chi_{n}-2\lambda_{n}\beta_{n}\right)\left\langle e_{x}^{2}e_{y}^{2}\right\rangle_{n}\Big]\label{eq:sigma_s}
\end{align}
The total volume of the aggregate is conserved in this approach $V_{ellipsoid}=nv_p$. The coefficients $\alpha_n$, $\beta_n$, $\lambda_n$, $\xi_n$, $\chi_n$ are the geometrical parameters of a chain\cite{Pokrovskii_1972,Zubarev_Iskakova_2000} and are given by eqs.~\eqref{eq:Ssigma_coeffs1}-\eqref{eq:Ssigma_coeffs2} (Supplemental Materials). The moments $<\dots>_n$ are calculated with respect to the non-equilibrium orientational distribution function $\phi_{n}\left(\bm{e}\right)$, which is the solution of the corresponding Fokker-Planck equation. The general solution is impossible and the perturbative solution\cite{Zubarev_Iskakova_1995,Zubarev_Iskakova_2000} has not yet been presented for strong fields. Assuming that $\phi_{n}\left(\bm{e}\right)$ is peaked around the deviation angle  $\theta_{n}$ \eqref{eq:tantheta_Mn}, justified by the considerations of mechanical stability, the moments can be roughly estimated $\left\langle e_{x}^{2} \right\rangle_{n}=\sin^2\theta_n$, $\left\langle e_{y}^{2} \right\rangle_{n}=\cos^2\theta_n$ and $\left\langle e_{x}^{2}e_{y}^{2} \right\rangle_{n}=\sin^2\theta_n\cos^2\theta_n$.
From eq.~\eqref{eq:sigma_a} one can show that the contribution of the $n$-mer to the antisymmetric stress scales as the length cubed $\propto n^{3}$, whereas for the symmetric stress \eqref{eq:sigma_s} the calculation is slightly more involved and yields $\propto\frac{n^{3}}{\ln{n}}$.

We have conducted simulated rheometric tests of the colloid in strong applied field and measured all components of the virial stress tensor in simple shear
\begin{equation}
 \sigma_{ij}=-\frac{1}{V}\sum_{k}{\left[ m^{\left(k\right)}\tilde{u}_{i}^{\left(k\right)}\tilde{u}_{j}^{\left(k\right)}-0.5\sideset{}{^{*}}\sum_{l}{\left(r_{i}^{\left(l\right)}-r_{i}^{\left(k\right)}\right)f_{j}^{kl}}\right]}+\sigma^{'}_{ij} \label{eq:virial_stress}
\end{equation}
where $\bm{\tilde{u}^{\left(k\right)}}=\bm{u^{\left(k\right)}}-\bm{\bar{u}}$ is the thermal velocity of a colloidal/MPCD particle $k$ relative to the measured average velocity field $\bm{\bar{u}}$ and $\sigma^{'}_{ij}$ is the component of the stress tensor peculiar to the MPCD solvent\cite{Winkler_Huang_2009}, which is associated with the redistribution of momentum in multi-particle collisions $\sigma^{'}_{ij}=-\frac{1}{V \Delta t_{mpcd}}\sideset{}{^{'}}\sum_{k}{r_{i}^{\left(k\right)}\Delta p^{\left(k\right)}_{j}}$, where $\Delta p^{\left(k\right)}=m\left(\bm{u^{\left(k\right)}}-\bm{\hat{u}^{\left(k\right)}}\right)$ and $\bm{\hat{u}^{\left(k\right)}}$ is the velocity before the collision ($\Sigma^*$ is taken just over the colloids and $\Sigma^{'}$- over the MPCD particles; $\Sigma=\Sigma^{'}+\Sigma^{*}$).

\begin{figure}[!h]
\centering
  \includegraphics[width=0.8\linewidth]{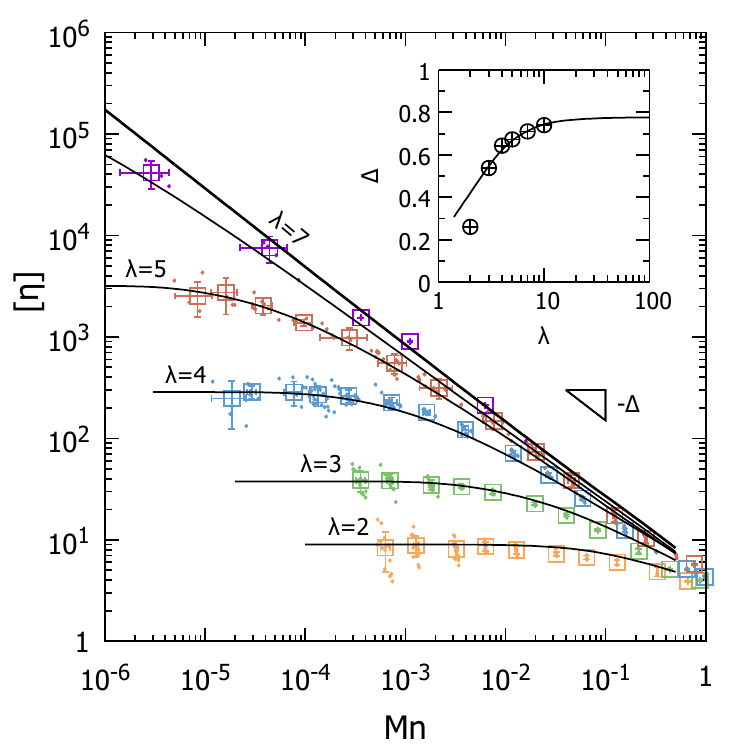}
  \caption{Measured intrinsic viscosity $\left[\eta\right]$ \eqref{eq:intrinsic} as a function of the reduced shear rate expressed in terms of the Mason number $Mn$ for various $\lambda=2,3,4,5,7$: simulation data ($\Box$), predictions of the chain model for effective dipole interaction parameter $\lambda_{*}=1.4,2.3,3.4,4.65,7$ (solid line). The bold line shows the limiting behavior $\lambda\gg 1$ predicted by the chain model. Inset: shear thinning exponent $\Delta$ ($\left[\eta\right]\propto Mn^{-\Delta}$) as a function of $\lambda$ calculated from the simulation data ($\oplus$) and the chain model (solid line).}
  \label{fgr:infres}
\end{figure}

The rheometric curves $\sigma_{xy}\left(\dot{\gamma}\right)$ and the viscosity $\eta=\frac{\sigma_{xy}}{\dot{\gamma}}$ in simple shear are the properties that are most frequently measured in a laboratory experiment. 
In turn, the intrinsic viscosity is defined as
\begin{equation}
\left[\eta\right]=\phi^{-1}\left( \frac{\sigma_{xy}}{\eta_{0}\dot{\gamma}}-1 \right)\label{eq:intrinsic}
\end{equation}
which is plotted in Figure~\ref{fgr:infres}. 

At high interaction parameters $\lambda\gg 1$ the intrinsic viscosity is a universal function of the Mason number $\left[\eta\right]\propto Mn^{-\Delta}$, where the exponent $\Delta$ becomes independent of $\lambda$. Here, the Mason number produces a gradual collapse of the shear thinning region \cite{Klingenberg_Ulicny_Golden_2007} as $\lambda$ increases. 
In turn, for intermediate $\lambda$ the Mason number alone is not sufficient to collapse the rheological response onto a single curve as it does not include the thermal effects and the exponent $\Delta\left(\lambda\right)$ becomes a function of $\lambda$, which leads to the branching of the viscometric curves. 
For low to moderate bonding energies $\lambda$ the calculated viscometric curves show at the lowest shear rates $Mn$ a pseudo-Newtonian plateau with a shear rate independent viscosity. 
At the higher shear rates a rapid shear thinning is observed, which is governed by a near power-law decrease of the intrinsic viscosity. For higher bond energies the power-law regions of the viscometric curves begin to converge, whereas the low shear plateau is elevated at approximately an exponential rate at the same time gradually moving to lower shear rates. At the largest $\lambda$ the plateau is outside the range of the Mason numbers accessible to the simulation and can no longer be resolved. 
The estimated shear thinning exponent $\Delta$ increases from about $\Delta\approx 0.3$ (at $\lambda=2$) to $\Delta\approx 0.7$ at $\lambda=7$. The limiting value $\Delta=1$ \citep{Martin_Anderson_1996} is not reached here. The model of Zubarev and Iskakova \citep{Yu_Zubarev_Yu_Iskakova_2007} explains the deviations from the reciprocal scaling with $Mn$, assuming that the chains are not identical and taking into account the polydispersity of the chain ensemble. The shear thinning exponents $\Delta<1$ are in accord with the experimental observations \citep{Berli_de_Vicente_2012} and Stokesian dynamics simulations in a monolayer \cite{Baxter_Drayton_Brady_1996}.

\subsection{Normal stresses}
\label{sec:normal_stresses}

Considering the WCA \eqref{eq:WCA} and dipole-dipole \eqref{eq:DD} interaction potential as a dissipative oscillator, the oscillatory motion of the particles is well damped by the viscous drag and the particles are mostly located at the bottom of the potential minimum, where the attraction force is compensated by steric repulsion. The thermal fluctuations introduce some tensile stresses even in equilibrium. A simple order of magnitude estimate shows that the kinetic contributions of the colloids to the virial stress are negligible. From eq.~\eqref{eq:virial_stress}
\begin{equation}
\sigma_{ii}=\frac{1}{2V}\sum_{k}\sum_{l}\left(r_{i}^{\left(l\right)}-r_{i}^{\left(k\right)}\right)f_{i}^{\left(kl\right)}
\label{eq:norm_sigma}
\end{equation}
Assuming that a net repulsive force $F_{r}=\frac{2\lambda kT}{\sigma}\bar{f}$ exists between each pair of particles in the chain aligned by a strong external field, the first normal stress difference can be straightforwardly estimated from eq.~\eqref{eq:norm_sigma} for a rigid chain as
\begin{equation}
N_{1}=\sigma_{xx}-\sigma_{yy}=\frac{6\bar{f}}{Mn}\eta_{0}\dot{\gamma}\sum_{n=2}^{n_{max}}\left( n-1 \right)\cos(2\theta_{n})g_{n}v_{p}
\end{equation}
where $\bar{f}$ is defined by the interaction potential. Assuming $\cos 2\theta_n\approx 1$
\begin{equation}
\left[N_1\right]=\phi^{-1}\frac{N_{1}}{\eta_{0}\dot{\gamma}}\approx \frac{6\bar{f}}{Mn}\left(1-\frac{1}{\left\langle n\right\rangle}\right)
\label{eq:norm_N1}
\end{equation}
which evidences that the first normal stress difference is, in fact, positive and a property of the bond and, hence, should be more or less independent of the shear rate, except for very large shear rates, when most of the larger bonded aggregates are eroded and the number of free monomers quickly increases $\left\langle n\right\rangle\rightarrow 1$.

\begin{figure}[h!]
\centering
    \includegraphics[width=0.7\linewidth]{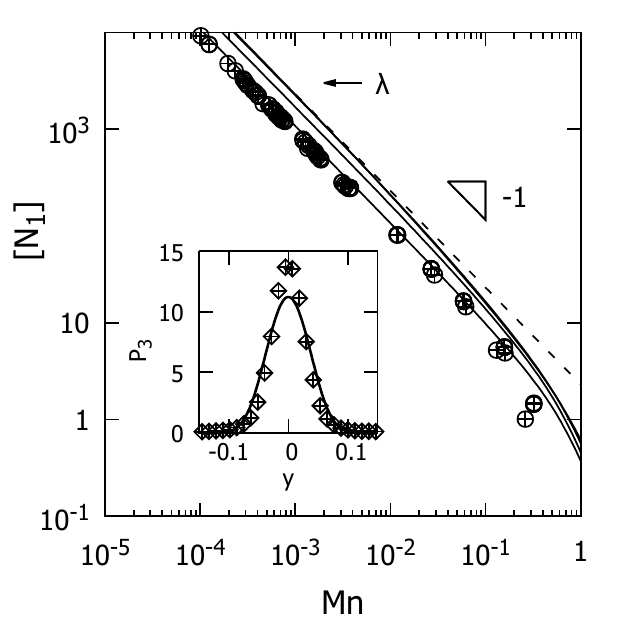}
  \caption{Normalized 1\textsuperscript{st} normal stress difference  $\left[N_1\right]$ calculated from simulations at $\lambda=2,3,4,5$ and $\phi=5\%$ ($\oplus$) and according to the predictions of the chain model \eqref{eq:norm_N1} for effective dipole interaction parameter $\lambda=0.82,1.55,2.483,3.484$ (solid lines). The simulation data does not depend on $\lambda$, the theoretical predictions only weakly depend on $\lambda$. Dashed line is $\propto Mn^{-1}$. Inset: equilibrium 1D positional probability density distribution function $P_3\left(y\right)$ of the central particle in a trimer along the director of the chain in strong external field $B_y$ at $\lambda=4$ ($\phi=5\%$): calculated from from simulations as $P_3\left(y\right)=\int{P_3\left(\bm{r}\right)}dxdz$ ($\diamondplus$) and according to eq.~\eqref{eq:norm_prob} (solid line). }
  \label{fgr:normal_prob}
\end{figure}

The repulsive force in question is originated by the thermal fluctuations\citep{Martin_2000}. The added Brownian stress develops from the fluctuation induced strain of the chain structure. A trimer (i.e. three-particle chain) coupled to a Langevin environment is a minimal representative chain model if considering only the nearest neighbor interactions. The positional probability density distribution $P_3$ of the central particle in the trimer fluctuating along the director and bound by two neighbors is described by a stationary solution of the corresponding Fokker-Planck equation. In the vicinity of the center of the chain:
\begin{equation}
P_{3}\left(r\right)\propto e^{-\beta U\left(r\right)}
\label{eq:norm_prob}
\end{equation}
where $\beta U\left(r\right)$ is the potential energy of the central particle in a trimer.
The inset of Fig.~\ref{fgr:normal_prob} shows the probability density distribution of a trimer in equilibrium for $\lambda=4$ plotted along the axis connecting the particles. It can be seen that the reference particle is localized in the potential well and the probability practically becomes zero below the chosen clustering cutoff criterion $r_{c}=1.1$. We have also checked the positional PDF of an end particle in a dimer relative to its pair and, despite slight asymmetry, found it similar to $P_3$ in shape and amplitude.
The particles in a chain oscillate between repulsive and attractive force. While both contributions on average compensate in a symmetric potential, they do not in an anharmonic potential, e.g. the dipole-dipole interaction coupled with a steric repulsion
\begin{equation}
\bar{f}=\frac{1}{2\lambda}\int-\frac{\partial\left( \beta U\right)}{\partial r} P_3\left(r\right) dr 
\label{eq:norm_fbar}
\end{equation}

The results (Figure~\ref{fgr:normal_prob}) show that the estimate \eqref{eq:norm_N1} over-predicts the normal stresses calculated from the simulation data, which is not surprising, considering the strength of the approximations. The second normal stress difference follows naturally $N_2=-N_1$. This is observed in the simulations within several percent (not shown here).
Estimates show that the fluctuation induced stress survives the hard sphere limit of eq.~\eqref{eq:WCA} $n\rightarrow\infty$ with $\bar{f}\propto n$.

\section{Experiments}
\label{sec:experiments}
\subsection{Ferrofluids}
\label{sec:ferrofluids}
\subsubsection{Dimensional analysis}
\label{sec:dimensional_analysis}
Magnetic colloids are inherently polydisperse systems, hence, the size variation translates ($\propto \sigma^3$) into a variation of the interaction parameter $\lambda$. The fraction of particles that are not large enough to form any structure $\lambda<1$ and exist predominantly in a non-aggregated state can be substantial. We assume a bidisperse structure of magnetic colloids, where the chains are primarily composed of larger particles $\lambda>1$ determining the rheological specificity \cite{M3,M5} and the smaller particles $\lambda<1$ are disregarded. Subsequently, we analyze the particle size distribution (PSD) of the investigated materials and renormalize it, discarding the influence of the smallest fraction. The estimated fraction of the large interacting particles is shown in column 10 of Table~\ref{tbl:exp_visc}. The average diameter is determined by averaging ($\propto\phi$) with respect to the remaining large fractions. If the normalization of the PSD is not performed to display the influence of the aggregating fraction, the precise position of the experimental curve on the $\left(Mn,\left[\eta\right]\right)$ plot cannot be obtained. Still, this is a rather crude approximation, since the small particles can participate in the interactions with the bigger particles and be included in the chains \cite{Ivanov_Kantorovich_2004,Holm_Ivanov_Kantorovich_Pyanzina_Reznikov_2006}, thereby influencing their length \cite{Zubarev_Iskakova_2004b}. This may significantly impact the rheological response due to the lower tensile strength of the induced defects or otherwise \cite{M26}.

\begin{figure}[!h]
\centering
  \includegraphics[width=0.8\linewidth]{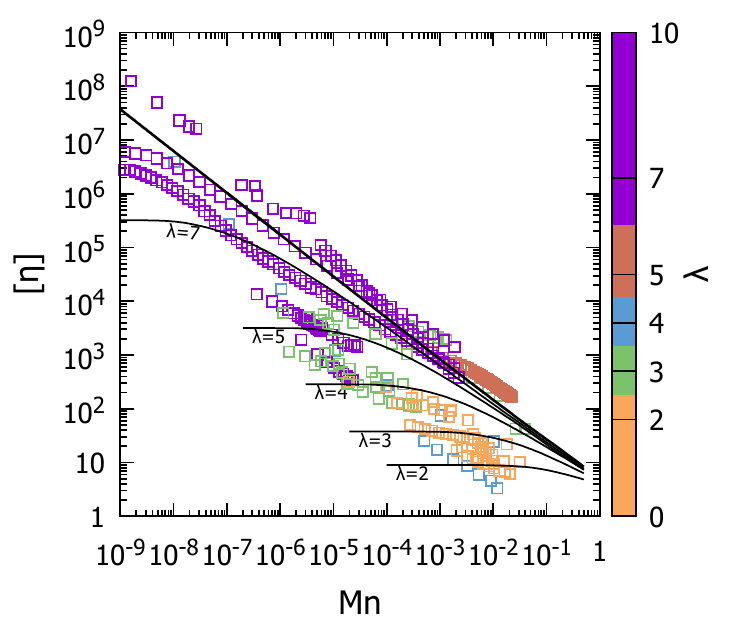}
  \caption{The experimental viscometric curves reported in the literature sources summarized in Table~\ref{tbl:exp_visc}, plotted in the coordinates of reduced shear rate (Mason number $Mn$) and intrinsic viscosity $\left[\eta\right]$. The color code corresponds to the estimated dipole interaction parameter $\lambda$. The solid lines indicate the chain model fit to the simulation data and are the same as in Figure~\ref{fgr:infres}. The bold line represents the limiting behavior $\lambda\gg 1$ of the chain model. At this point the comparison allows to establish solely the order-of-magnitude correspondence of the experimental and simulation data.}
  \label{fgr:exp_visc}
\end{figure}

An additional difficulty comes from the fact that only some of the published investigations report all the relevant material parameters necessary for modeling and quantitative comparison. The magnetometric measurements\citep{ws1985} are useful to characterize the magnetic size $\sigma_m$ distribution and magnetic moment $\mu=\frac{1}{6}\sigma_m^3 M_s$ (here $M_s$ is the spontaneous magnetization of the particle material) distribution of the spontaneously polarized particles \citep{M1,M15,M16,M27,M28} and estimate the dipole-dipole interaction parameter $\lambda$. One must take into account that the size of the magnetic core can be significantly smaller than the hydrodynamic diameter. The difference is associated with the existence of a stabilizing shell (e.g. a layer of surfactant) covering the monodomain as well as the magnetically "dead" layer on its surface. The total size of the solid grain can be estimated by transmission electron microscopy (TEM) \citep{M26,M23}. In turn, the dynamic light scattering (DLS)\citep{M23,M24} can be used to determine the hydrodynamic diameter directly. An exhaustive characterization of the ferrofluid using different methods facilitates the determination of $\sigma_m$ and $\sigma$. 
The hydrodynamic volume fraction $\phi$ contained in \eqref{eq:intrinsic} is an awkward parameter, which is almost never reported. It is assumed that $\phi=\frac{\sigma^3}{\sigma_m^3}\phi_m$, where $\phi_m$ is the volume fraction of the dispersed magnetic material, which is usually 2-3 times smaller than $\phi$.
The summary of the previously conducted rheological investigations of ferrofluids is shown in Table~\ref{tbl:exp_visc}, which is more or less self-explanatory, and the results are presented in Figure~\ref{fgr:exp_visc} in a $Mn-\left[\eta\right]$ plot. In the analysis we use only the measurements at the highest reported magnitudes of the applied field, thereby assuming that the investigated ferrofluids are more or less at saturation, which may not always be completely true.

\begin{table*}
\small
  \caption{A summary of some of the previous rheological investigations of magnetic colloids (ferrofluids) from various literature sources: reference, designation, type of magnetic material and protective shell, estimated diameter of the magnetic core $d_m$ of the interacting fraction, estimated particle hydrodynamic diameter $D_h$ (distance of closest approach $\sigma$) including the stabilizing layer, type of solvent, estimated viscosity $\eta_0$ of the pure solvent, volume fraction of the dispersed magnetic material $\phi_m$ (total and estimating the amount of the interacting particles $\lambda\geq 1$), investigated range of shear rates $\dot{\gamma}$, highest investigated field $B$, estimated dipole interaction parameter $\lambda$ }
  \setlength{\tabcolsep}{0.65em}
  \label{tbl:exp_visc}
  \begin{tabular*}{\textwidth}{@{\extracolsep{\fill}}rlllrrlrrrrrrll}
    \hline
         &           & \multicolumn{4}{c}{Dispersed phase}                                     & \multicolumn{2}{c}{Medium} &  \multicolumn{2}{c}{$\phi_{m}$, vol\%} &           &      & \\
    Ref.             & Sample     & Material         & Shell             & $d_{m}$, nm & $D_{h}$, nm & Solvent & $\eta$, mPa & Total & $\lambda\geq 1$ & $\dot{\gamma}$, s$^{-1}$ & B, mT & $\lambda$\\
    \hline
    \citep{M1}     & APG513A    & Fe$_3$O$_4$      & \textit{unknown}       & $14$ & $20$ & ester            & $120$  & $6.7$   & $0.6$  & $ 1-5$       & $10$ & $1.3$\\
    \citep{M2}     & APG513A    & Fe$_3$O$_4$      & \textit{unknown}       & $14$ & $20$ & ester            & $120$  & $7.2$   & $0.6$  & $15-37$      & $40$ & $1.3$\\
    \citep{M5}     & APG513A    & Fe$_3$O$_4$      & \textit{unknown}       & $16$ & $20$ & ester            & $120$  & $7.2$   & $0.7$  & $0.05-2$     & $30$ & $2.7$\\
    \citep{M4,M6}  & APG513A    & Fe$_3$O$_4$      & \textit{unknown}       & $17$ & $21$ & ester            & $120$  & $7.0$   & $0.5$  & $0.1-1$      & $40$ & $3.2$\\
    \citep{M6}     &            & Cobalt           & Al$_2$O$_3$+KorantinSH & $10$ & $16$ & L9 oil           & $ 65$  & $0.4$   & $0.4$  & $0.05-1$     & $40$ & $3.5$\\
    \citep{M7}     & APG513A    & Fe$_3$O$_4$      & \textit{unknown}       & $16$ & $20$ & ester            & $120$  & $7.2$   & $0.7$  & $0.05-2$     & $50$ & $2.7$\\
    \citep{M7}     & APG513A    & Fe$_3$O$_4$      & \textit{unknown}       & $16$ & $20$ & ester            & $120$  & $7.2$   & $0.7$  & $0.05-2$     & $50$ & $3.2$\\
    \citep{M9}     & APG513A    & Fe$_3$O$_4$      & \textit{unknown}       & $16$ & $20$ & ester            & $120$  & $7.2$   & $0.7$  & $0.05-2$     & $150$ & $2.7$\\
    \citep{M9}     & APG513A    & Fe$_3$O$_4$      & \textit{unknown}       & $16$ & $20$ & ester            & $120$  & $7.2$   & $0.7$  & $0.05-2$     & $150$ & $2.5$\\

    \citep{M9,M10}& Co87       & Cobalt           & Al$_2$O$_3$+KorantinSH & $10$ & $16$ & kerosene         & $2.5$  & $0.35$  & $0.35$ & $1-400$      & $130$ & $3.4$\\    
    \citep{M15}   & Co87       & Cobalt           & Al$_2$O$_3$+KorantinSH & $13$ & $19$ & kerosene         & $2.5$  & $0.21$  & $0.16$ & $1-8$        & $30$ & $11$\\
    \citep{M16} & S2         & Cobalt           & \textit{unknown}       & $12$ & $18$ & kerosene         & $2.5$  & $0.13$  & $0.11$ & $0.5-8$      & $40$ & $8.4$\\
    \citep{M20}   & NS FF      & CoNi             & L-$\alpha$-PC          & $24$ & $30$ & mineral oil      & $39.6$ & $5.1$   & $5.1$  & $10^{-5}-175$& $440$ & $29$\\
    \citep{M22}   & FF $\phi$=0.038 & Fe$_3$O$_4$ & oleic acid             & $19$ & $23$ & kerosene         & $2.5$  & $3.8$   & $2.2$  & $10^{-6}-10^{3}$ & $200$  & $6.7$\\
    \citep{M23}   & FF-A       & \textit{unknown} & \textit{unknown}       & $16$ & $20$ & \textit{unknown} & $9  $  &   $8$   & $4.2$  & $10^{2}-10^{3}$  & $280$ & $1.7$\\
    \citep{M23}   & FF-B       & \textit{unknown} & \textit{unknown}       & $16$ & $20$ & \textit{unknown} & $6  $  &   $7$   & $7$    & $1-10^{3}$       & $280$ & $2.5$\\
    \citep{M24}   & FF         & Fe$_3$O$_4$      & oleic acid             & $19$ & $23$ & kerosene         & $2.5$  & $8.8$   & $4.1$  & $10^{-4}-10^{3}$ & $340$ & $6.4$\\

    \citep{M26}    & A0         & CoFe$_2$O$_4$    & citric acid         & $35$ & $40$ & water            & $1  $  & $0.23$  & $0.23$ & $3-27$    & $40$ & $28$\\
    \citep{M26}    & A0.75      & CoFe$_2$O$_4$    & citric acid         & $28$ & $33$ & water            & $1  $  & $0.91$  & $0.39$ & $3-27$    & $40$ & $13$\\
    \citep{M27}   &            & Fe$_3$O$_4$      & \textit{unknown}       & $17$ & $21$ & ester            & $34 $  & $6.3$   & $3.1$  & $0.5-10$  & $50$ & $3.2$\\
    \citep{M28}   & Co87       & Cobalt           & Al$_2$O$_3$+KorantinSH & $16$ & $22$ & kerosene         & $2  $  & $0.72$  & $0.66$ & $0.5-10$  & $60$ & $22$\\
    \citep{M29}   & MagSilica  & Fe$_3$O$_4$/Fe$_2$O$_3$ & SiO$_{2}$+C8TMS & $18$ & $26$ & AP201 oil        & $68 $  & $1.3$   & $1.3$  & $0.1-100$ & $40$ & $2.5$\\
    \citep{M30}   & Sample 5   & Fe$_3$O$_4$      & OA/SA/ASA              & $19$ & $23$ & PES-5            & $250$  & $3.6$   & $1.6$  & $1-100$   & $130$ & $5.1$\\
    \hline
  \end{tabular*}
\end{table*}

The experimentally measured viscosities of the various ferrofluids span 8 orders of magnitude across 8 orders of magnitude of the shear rate at quite moderate applied fields. Although some trends are already visible - e.g. increase of the viscosity with a larger dipolar interaction parameter - the interpretation of this chart is not straightforward due to the fact that the measurements were performed at different effective temperatures, i.e. dispersion in $\lambda$. The Mason number alone is not a reliable scaling parameter for ferrofluids and it is not sufficient to collapse the viscometric data. However, the collapse can be achieved with additional processing.

\subsubsection{Viscosity master curve}
\label{sec:viscosity_master_curve}
The time-temperature superposition\cite{ferry1980viscoelastic} is performed to produce the viscosity master curve. The construction is achieved using the shift factors $\alpha_{\eta}$ and $\beta_{\eta}$ and is based on the observed analogy between thermal and hydrodynamic erosive stresses, i.e. the increase of the reduced temperature $T^{*}$ (or decrease of the $\lambda$ parameter) bears a similar effect on the viscoelactic properties of magnetic colloids as the increase of the shear rate expressed in terms of the Mason number Mn. Hence, the isothermal viscometric curves obtained at different temperatures (or $\lambda$) over a range of shear rates can be superposed by a scaling relation.

From eq.~\eqref{eq:sigma_a} and accounting for eqs.~\eqref{eq:gn_theor} and \eqref{eq:X} it follows that for vanishing shear rates $Mn\rightarrow 0$: $\frac{\sigma^{a}}{\eta_{0}\dot{\gamma}}=0.75 e^{-\epsilon_{*}}\left[Li_{-3}\left(X\right)-Li_{-1}\left(X\right)\right]$, where $Li_n\left(x\right)$ is a polylogarithm. For larger dipolar strengths $\lambda\gg 1$:
\begin{equation}
\phi^{-1}\frac{\sigma^{a}}{\eta_{0}\dot{\gamma}}\approx4.5\phi e^{\epsilon_{*}} \label{eq:sa_plateau}
\end{equation}
In the chain model the low shear rate plateau is dominated by the antisymmetric stress, so equation \eqref{eq:sa_plateau} is a good approximation for eq.~\eqref{eq:intrinsic} at $Mn\rightarrow 0$. It also shows that the intrinsic viscosity is not independent of concentration, which is appropriate for self-assembled systems, and scales exponentially with the binding energy. Equation \eqref{eq:sa_plateau} is formally in accord with the Arrhenius model where $\epsilon_{*}$ carries the meaning of the activation energy. The vertical shift factor is defined using the low shear viscosity as a scaling parameter $\beta_{\eta}=\left[\eta\right]_{0}^{-1}$.

In turn, the horizontal shift factor $\alpha_{\eta}=Mn_{*}^{-1}$ is defined in terms of the critical Mason number $Mn_{*}$, i.e. the point (horizontal intercept), where the straight lines corresponding to the power-law segment $\propto Mn^{-\Delta}$ and the plateaus of the viscometric curves intersect on a log-log plot, which approximately marks a transition from the pseudo-Newtonian plateau, dominated by dipolar interactions, into the shear-thinning regime. Both shift factors scale exponentially with the dipole-dipole interaction parameter at $\lambda\gg 1$, whereas the small deviations at low to moderate values of $\lambda$ are the effect of the Brownian motion and the differences between the nominal and effective values of the interaction parameter discussed previously. It is assumed that for $\lambda\geq 7$ the value of the effective dipole-dipole interaction parameter is the same as the nominal $\lambda_{*}=\lambda$.

\begin{figure}[!h]
\centering
  \includegraphics[width=0.8\linewidth]{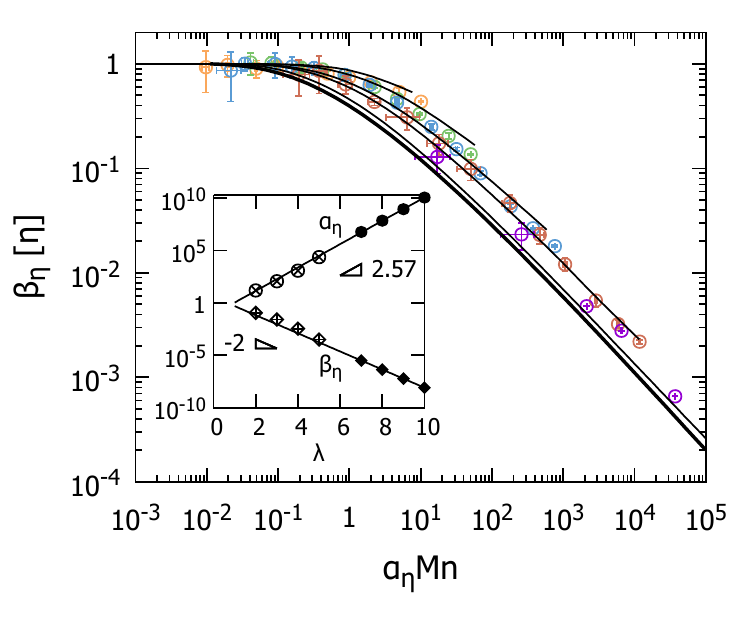}
  \caption{Simulation data of Figure~\ref{fgr:infres} plotted in scaled coordinates $\left(\alpha_{\eta}Mn,\beta_{\eta}\left[\eta\right]\right)$. The bold line represents the limiting behavior $\lambda\gg 1$ of the chain model. The inset shows the scaling coefficients $\alpha_{\eta}$ and $\beta_{\eta}$ as a function of $\lambda$: obtained by collapsing the simulation data ($\otimes$, $\diamondplus$) and theoretical curves predicted by the chain model for $\lambda\geq 7$ ($\bullet$, $\filleddiamond$). The solid lines are power law fits $\alpha_{\eta}=0.0825e^{2.57\lambda}$ and $\beta_{\eta}=3.58e^{-2\lambda}$.}
  \label{fgr:MCvisc}
\end{figure}

In Figure~\ref{fgr:MCvisc} we plot the simulation data using the time-temperature superposition in the normalized coordinates $\left(\alpha_{\eta}Mn,\beta_{\eta}\left[\eta\right]\right)$. As the interaction parameter is increased $\lambda\gg 1$, we see evidence of a limiting behavior, where both the data and the theoretical predictions by the chain model converge onto a single universal master curve. In the past a similar construction was tried by Baxter-Drayton et al.\cite{Baxter_Drayton_Brady_1996} based on purely empirical considerations to interpret their monolayer simulations of electrorheological fluids.
\begin{figure}[!h]
\centering
  \includegraphics[width=0.8\linewidth]{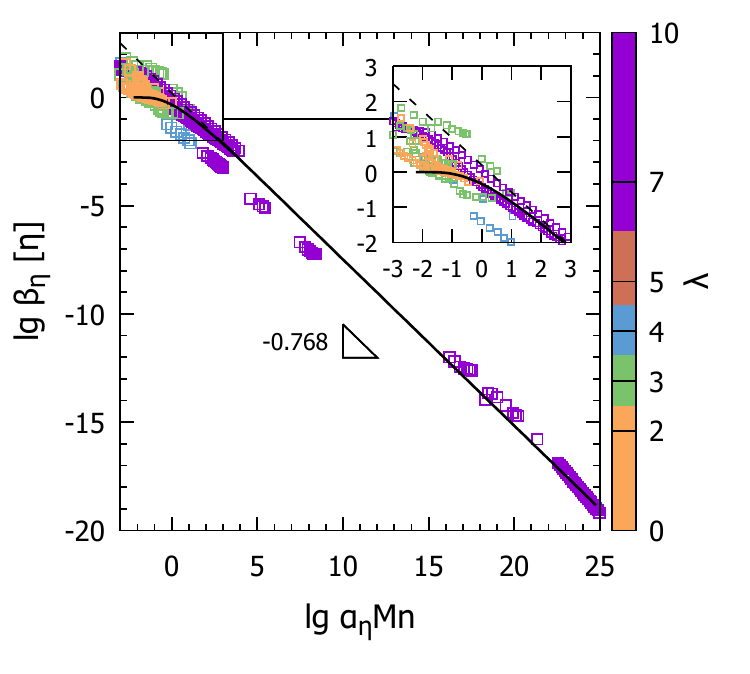}
  \caption{Experimental data of Figure~\ref{fgr:exp_visc} re-plotted in scaled coordinates $\left(0.0825e^{2.57\lambda}Mn,3.58e^{-2\lambda}\left[\eta\right]\right)$. The bold line represents the limiting behavior $\lambda\gg 1$ of the chain model. The dashed line is a power law $1.57 x^{-0.768} $.}
  \label{fgr:expMC}
\end{figure}

Next, we re-plot the experimental data of Figure~\ref{fgr:exp_visc} in the normalized coordinates $\left(\alpha_{\eta}Mn,\beta_{\eta}\left[\eta\right]\right)$, where $\alpha_{\eta}=0.0825e^{2.57\lambda}$ and $\beta_{\eta}=3.58e^{-2\lambda}$ are determined from the time-temperature superposition of simulation data and theoretical predictions at our standard volume fraction $\phi=5\%$. Inspecting Figure~\ref{fgr:expMC} we find robust experimental evidence of a universal behavior predicted by the simulations and the chain model. By the use of this procedure the dispersion in $\lambda$ associated with the varying material properties (Table~\ref{tbl:exp_visc}) is collapsed to a significant degree. The benefit of this scaling is that it shifts the experimental window, broadening the range of the accessible effective shear rates in terms of the Mason number $Mn$, which allows to examine greater portions of the master curve using different materials, which are not accessible with a single ferrofluid.
There are several remaining sources of dispersion in the mapping of the experimental results: 
\begin{enumerate*}[label=(\alph*)]
\item the uncontrolled contribution to the interaction potential comes from the residual short-range isotropic attraction, e.g. leaking through the steric barrier or due to the defects in the protective shell. The Van der Waals interaction is of the order $\propto kT$ and is likely significant in some weakly-associating ferrofluids $\lambda\approx1-3$ based on synthetic magnetite (Fe\textsubscript{3}O\textsubscript{4}) \cite{Butter_Bomans_Frederik_Vroege_Philipse_2003a}
 
\item difficulty of obtaining saturation with respect to the external field in some measuring arrangements, e.g. the group of data\cite{M4,M15,M16,M26} directly underneath the master curve in Figure~\ref{fgr:expMC} was measured at $30-40$~mT, which is apparently not at saturation of the magnetic viscosity effect, hence, they display the same dependence, but do not quite reach the predicted magnitude 

\item incomplete compensation of the effect of polydispersity and the dispersion of the interaction parameter, especially at low values of $\lambda$. Similarly, incomplete characterization of the ferrofluids in general, requiring assumptions.

\item the independence of the intrinsic viscosity on concentration is not completely established (Figure~\ref{fgr:Svisc_conc}, Supplemental Materials) in the vicinity of the plateau, whereas the concentration of the chain-forming particles varies across at least one order of magnitude (Table~\ref{tbl:exp_visc}) in the investigated samples.

\end{enumerate*}. 

There are many other uncontrolled parameters, which can significantly influence the viscoelastic properties of the magnetic colloid. For one, the amount of free surfactant or primary agglomerates \cite{Buzmakov_Pshenichnikov_1996} remaining in the solution from the preparation stage is difficult to quantify, but it can significantly influence the measured viscosity.

\subsection{Other dipolar systems}
\label{sec:other_dipolar_systems}

\begin{figure}[!h]
\centering
  \includegraphics[width=0.7\linewidth]{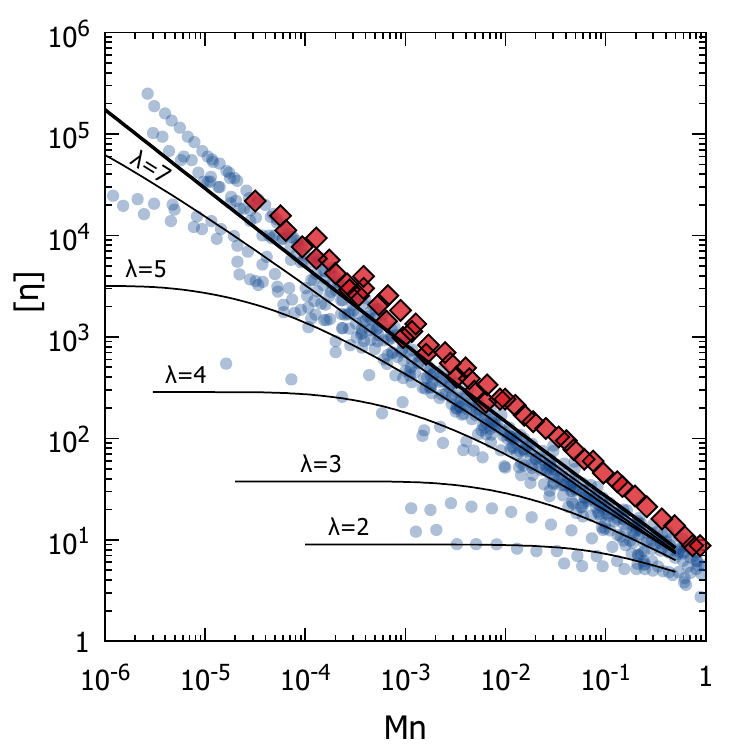}
  \caption{Experimental viscometric curves of "inverse ferrofluids" ($\circ$) and superparamagnetic colloids ($\diamond$), reported in the literature sources (see Section~\ref{sec:other_dipolar_systems} for details), plotted in the coordinates of reduced shear rate (Mason number $Mn$) and intrinsic viscosity $\left[\eta\right]$. The solid lines indicate the chain model fit to the simulation data and are the same as Fig.~\ref{fgr:infres} and \ref{fgr:exp_visc}. }
  \label{fgr:other_visc}
\end{figure}

The interpretation of the experimental data for ferrofluids is challenging because of the lack of a well-defined model system and the sensitivity of their structural and rheological properties to various parameters, which cannot be controlled or accurately measured. In turn, the dipolar systems, where the magnitude of the interactions can be varied, have proven to be excellent experimental benchmarks to test the theoretical models of dipolar materials\cite{Berli_de_Vicente_2012,Ruiz-Lopez_Fernandez-Toledano_Hidalgo-Alvarez_deVicente_2016,Ruiz-Lopez_Fernandez-Toledano_Klingenberg_Hidalgo-Alvarez_deVicente_2016}. The "inverse ferrofluids" (i.e. magnetic "holes") are obtained by dispersing inert particles in a ferrofluid matrix\cite{Skjeltorp_1983}, whereby a tunable permeability mismatch induced by the external field produces a virtual dipole associated with each "hole". The magnetorheology of inverse ferrofluids based on nonmagnetic silica nanoparticles (diameter $\sigma\approx100-700$nm) dispersed at a volume fraction $\phi\approx1-26\%$ has been previously reported as a function of the shear rate and magnetic field strength in terms of the Mason number\cite{ de_Gans_Hoekstra_Mellema_1999, deGans_Duin_vandenEnde_Mellema_2000,  Ramos_Klingenberg_Hidalgo-Alvarez_deVicente_2011, Ruiz-Lopez_Fernandez-Toledano_Klingenberg_Hidalgo-Alvarez_deVicente_2016}. The corresponding intrinsic viscosity is reproduced in Figure~\ref{fgr:other_visc} along with the fits to our simulation data using the chain model showing overall good agreement.

Unlike the magnetic "holes", the superparamagnetic colloids\cite{Faraudo_Andreu_Calero_Camacho_2016} are magnetically inert microparticles impregnated with single-domain magnetic grains. Hence, they display a superparamagnetic response to the applied field. The viscoelastic behavior of the composite polystyrene spheres (diameter $\sigma\approx 1\mu m$) laden with maghemite (Fe\textsubscript{2}O\textsubscript{3}) and dispersed in an aqueous solution at a volume fraction $\phi= 5.2\%$ was previously reported\cite{Felt_Hagenbuchle_Liu_Richard_1996} as a function of the Mason number $Mn$ and is also plotted in Figure~\ref{fgr:other_visc}. The dipole interaction parameter of the superparamagnetic colloids is rather large $\lambda\gg 1$ owing to the quantity of contained magnetic material ($>67$wt.\%) and their response approaches the limiting behavior of the chain model.

\section{Phase diagram}
\label{sec:phase_diagram}
\begin{figure}[!h]
\centering
  \includegraphics[width=0.8\linewidth]{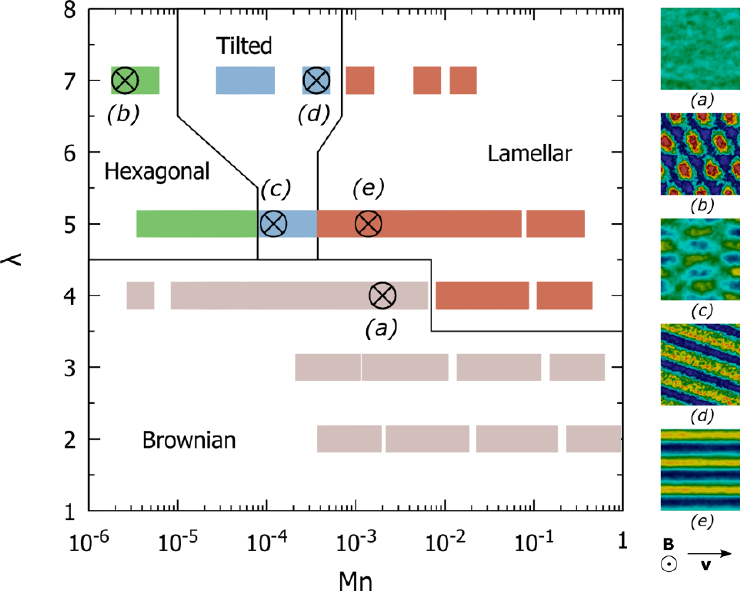}
  \caption{A tentative phase diagram of observed mesoscopic phases induced by strong field and shear flow at a volume fraction $\phi=5\%$: $\left(a\right)$ at lower interaction strength $\lambda\leq 4$ the Brownian phase is characterized by monomers randomly diffusing through all clusters in the system resulting in a homogeneous monomer spatial density distribution; $\left(b-e\right)$ at higher $\lambda\geq 5$ there are 3 distinct phases characterized by non-homogeneous monomer density patterns looking from the direction of the field onto the velocity-vorticity plane - $\left(b\right)$ affinely deformed hexagonal phase dominated by dipolar interactions, $\left(c\right)$ unstable fluctuating situation in the vicinity of the critical point ($\lambda_{c}\approx 5$, $Mn_{c}\approx 10^{-4}$) and shear-banded states $\left(d\right)$ - tilted lamellar (layered) phase, $\left(e\right)$ - aligned lamellar phase.  }
  \label{fgr:lamellar}
\end{figure}

The microstructural transformations determine the complicated viscoelastic response of dipolar colloids. 
In Figure~\ref{fgr:lamellar} a tentative phase diagram is presented showing various mesoscopic phases that we encountered in the simulations of the field-structured dipolar colloids under shear. 

\begin{figure}[!h]
\centering
  \includegraphics[width=0.925\linewidth]{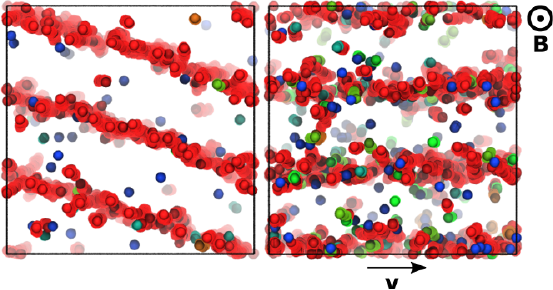}
  \caption{Simulation snapshots (box size $24\sigma\times 72\sigma\times 24\sigma$) of lamellar phases in simple shear and a strong vertical applied field $\bm{B}$ (viewed from the direction of the field) at a volume fraction $\phi=5\%$, corresponding to points $\left(d\right)$ and $\left(e\right)$ on the phase diagram (Fig.~\ref{fgr:lamellar}): left - tilted lamellar phase ($\lambda=7$ and $Mn=3.6\cdot 10^{-4}$), right - aligned lamellar phase ($\lambda=5$ and $Mn=1.4\cdot 10^{-3}$). Particles color-coded by aggregate size $n$: monomers ($n=1$) are blue, $n\geq10$ - red.  }
  \label{fgr:lamellar_top}
\end{figure}
At small $\lambda$ (Brownian phase) the particles are free to diffuse through all spontaneously forming aggregates and the lamellar phase does not form. In turn, in systems dominated by dipolar interactions $\lambda\geq 5$ the chains approach laterally, aggregating into thick chain-like bundles (Figure~\ref{fgr:l57_snaphots}), which can already be identified as persistent individual entities (rather than brief structural fluctuations). The bundles of chains form a hexagonal order (Figure~\ref{fgr:l57_snaphots}a,b) in the velocity-vorticity plane and the lateral mobility of the particles is restricted. Under simple shear the stability of the hexagonal phase is diminished and the perturbation of the dissolved microstructure exposes new pathways in the form of a phase of flowing layers of chains interleaved by slip planes of pure carrier liquid. The competition of the thermal, polarization and shearing stresses directs the associated mesophase transformations. Above the Brownian transition ($\lambda\geq 5$) at weak shear ($Mn<10^{-5}-10^{-4}$) the effect of the flow is limited to affine deformation of the chain bundles (Figure~\ref{fgr:lamellar}a). Increasing the shear rate the hexagonal pattern transforms into a tilted lamellar phase (Fig.~\ref{fgr:lamellar}d and Fig.~\ref{fgr:lamellar_top}, left), where the individual layers are oriented at approx. $\pm 20^{\circ}$ to the direction of the shear flow. Upon further increase of the shear rate the lamellas abruptly reorient in the direction of the shearing velocity at $Mn\approx 10^{-3}$ (Fig.~\ref{fgr:lamellar}e and Fig.~\ref{fgr:lamellar_top}, right). The period of the shear banded structure is $L=8\sigma$ for $\lambda=4-7$ and is independent of the shearing rate. This conventional lamellar phase can be directly accessed from the Brownian phase by increasing the strength of the dipolar interactions $\lambda$ at high shear rates $Mn>10^{-3}$. In the vicinity of the critical point ($\lambda_{c}\approx 5$, $Mn_{c}\approx 10^{-4}$), where the Brownian, lamellar and tilted lamellar phases meet, we observe unstable lamellar phases oscillating in the shear flow between $+20^{\circ}$ and $-20^{\circ}$, where the chain bundles are slipping relative to each other in a complicated hexagonal motion. This behavior is masked by the intense Brownian fluctuations of the particles and is only visible on the averaged monomer density snapshots of the vorticity-velocity plane (Figure~\ref{fgr:lamellar}c). We note that any delamination transition is marked by an increase of the critical chain length $n_{crit}$ (Figure~\ref{fgr:ncrit_navg}), evidencing that the transversal erosive stresses are significantly reduced. The previously calculated phase diagram \cite{Melrose_1991} in $Pe-\lambda$ space has only limited qualitative agreement with Figure~\ref{fgr:lamellar}, most likely due to the idealized hydrodynamics used in the previous works.  In particular, the existence of a tilted lamellar phase is a new development. Delaminated patterns were reported in an ER fluid under steady \cite{Henley_Filisko_1999, Henley_Filisko_2002, Cao_Huang_Zhou_2006, Li_Huang_Tang_Huang_Zhang_Zhou_2012} and oscillatory shear \cite{Bossis_Grasselli_Lemaire_Persello_Petit_1994}. In turn, suspensions of $0.5 \mu m$ magnetic polystyrene particles showed the onset of a regular mesoscopic striped structure above a critical shear rate \cite{Volkova_Cutillas_Bossis_1999}. The phase diagram (Figure~\ref{fgr:lamellar}) shows that the delamination transition is completely within the range of characteristic parameters of ferrofluids (Table~\ref{tbl:exp_visc}).

\section{Conclusions}
\label{sec:conclusions}
We have used hybrid simulations with coarse-grained hydrodynamics to observe the self-assembly of dipolar colloids with significant interparticle interactions $\lambda>1$ and determine their rheological behavior in simple shear across six orders of magnitude of the Mason number $Mn$ for volume fractions $\phi=1-9\%$. The emergence of various aligned structures is observed across different length scales in equilibrium and non-equilibrium scenarios. 
We show by simulation and theoretically that a master curve exists with respect to the chain-length distribution at moderate values of the interaction parameter $\lambda\leq 4$. 
Non-equilibrium simulations with imposed shear flow revealed conformational changes in the monomer PDF and a structure defined by the simultaneous competition of field induced alignment, hydrodynamic torque, erosive stresses and dipolar bond strength. 
We provide evidence of the universality of the structural behavior governed by the competition of bonding (dipolar) and dissociating (thermal and/or hydrodynamic) stresses.
The dynamical properties reflect the conformational transformations: for bond strengths $2\leq\lambda\leq 5$, the dipolar colloids display typical polymeric rheology, i.e. showing a pseudo-Newtonian plateau with constant viscosity for vanishing shear rates $Mn\rightarrow 0$ followed by a near-power law shear-thinning behavior $\left[\eta\right]\propto Mn^{-\Delta}$ at higher shear rates. For $\lambda>5$ the plateau is outside the simulation window resulting in an apparent yield stress and pseudo-plastic Bingham-like rheology. 
The simulations show that a meaningful description of the observed flow behavior can be obtained within a simple model of rigid rod-like chains with statistically distributed number of bonded particles. 
Quite surprisingly, despite several strong assumptions concerning the flexibility of the chains and disregarding chain-chain interactions (e.g. hydrodynamic), it provides a fair account of both the pseudo-Newtonian plateau and the shear thinning region of the viscometric curve. To ensure this, the model is calibrated to the simulation data using the dipolar interaction parameter $\lambda$ as a semi-empirical coefficient. We have discussed, which approximations should be improved to achieve better quantitative performance of the chain model. 

The actual self-assembled microstructure persists across different length scales and is more complex than that of rigid single particle chains: above a critical interaction parameter $\lambda\geq 5$ there is an abrupt shear-induced delamination transition leading to the formation of densified layered (lamellar) phases. Depending on the Mason number $Mn$ three distinctly different flow regimes are observed: at low shear, the hexagonally arranged self-assembled bundles of chains  support each other and resist the imposed stress, experiencing only affine deformations. Increasing the shear rate, the bundles start slipping, eventually leading to shear banding and forming of a tilted lamellar phase with layers of chains flowing at an angle $\pm 20^{\circ}$ to the direction of shear. At higher shearing rates the phases become fully aligned with the flow. The emergence of the layered phases affects the stress distribution in the system, which ultimately manifests itself in the lengthening of the chains.

In view of the used idealizations we obtain fair quantitative agreement between NEMD simulations and the viscoelastic response of magnetic colloids reported in various literature sources. This is achieved via the time-temperature superposition supported by the theoretical insight gained from the chain model and without using any fitting parameters. For the first time we have shown here evidence of a universal behavior of the rheology of ferrofluids obeying a single master curve. The observed collapse of the experimental viscometric curves substantiates the usefulness of the modeling approach. In fact, we provide evidence for the universality of the dipolar framework, which can be applied generally across very different length scales and physical systems.

Hopefully, this work will turn out to be a useful advancement towards developing a more general quantitative model of the rheology of dipolar colloids with a robust predictive strength for more complex geometries (incl. rheometric channels) and flows.

\section{Acknowledgments}

The financial support of Sciex-NMS\textsuperscript{ch} (Project 13.124) and the European Research Council (ERC) Advanced Grant 319968 FlowCCS is gratefully acknowledged. E.B. acknowledges the support of the LSR Program IMIS2.





\bibliography{rsc} 
\bibliographystyle{rsc} 

\pagebreak
\begin{center}
\textbf{\large Supplemental Materials: Self-assembly and rheology of dipolar colloids in simple shear - studied by multi-particle collision dynamics}
\end{center}
\setcounter{equation}{0}
\setcounter{figure}{0}
\setcounter{table}{0}
\setcounter{page}{1}
\makeatletter
\renewcommand{\theequation}{S\arabic{equation}}
\renewcommand{\thefigure}{S\arabic{figure}}
\renewcommand{\bibnumfmt}[1]{[S#1]}
\renewcommand{\citenumfont}[1]{S#1}

Accounting for just the nearest neighbor interactions the partition function of a self-assembled chain is \citep{S_Osipov_Teixeira_TelodaGama_1996}
\begin{equation}
  Z_n = \int e^{-\beta\sum_{i=1}^{n-1}U_{dd}\left(\bm{\mu_i},\bm{\mu_{i+1}},\bm{r_{i,i+1}}\right)+U_{ss}\left(r_{i,i+1}\right)}\prod_{i=1}^{n}\frac{\bm{d\mu_i}}{4\pi}\prod_{i=1}^{n-1}\bm{dr_{i,i+1}} \label{eq:SZn}
\end{equation}
whereas the interactions beyond the directly adjacent particles along the chain lead to the renormalization of the bond strength $\lambda \rightarrow \lambda \zeta\left(3\right), \zeta\left(3\right)=1.202$. \citep{S_Jordan_1973, S_Morozov_Shliomis_2004}
The pair partition function $Z_2$ in zero and infinite field \citep{S_Morozov_Shliomis_2004}
\begin{gather}
  H\rightarrow 0: Z_2\left(0\right) = 8\lambda \int_{0.5}^{\lambda}\frac{dl}{l^3}\int_0^1 dx \frac{\sinh\left(l\frac{1-3x^2}{2}\right)}{1-3x^2}I_0\left(3l\frac{1-x^2}{2}\right) \label{eq:SZ2zero} \\
  H\rightarrow \infty: Z_2\left(\infty\right) = 4\lambda \int_{0.5}^{\lambda}\frac{dl}{l^2}\int_0^1 dx   e^{l\left(3x^2-1\right)} \label{eq:SZ2infty}
\end{gather}
The geometrical coefficients used in eq.~\eqref{eq:sigma_s} are\cite{S_Pokrovskii_1972,S_Zubarev_Iskakova_2000,S_Yu_Zubarev_Yu_Iskakova_2007}
\begin{gather}
\alpha_n = \frac{1}{n\alpha_0^\prime},\quad \beta_n = \frac{2\left(n^2-1\right)}{n\left(n^2\alpha_0+\beta_0\right)},\quad \xi_n = \frac{4}{\left(n^2+1\right)n\beta_0^\prime}-\frac{2}{n\alpha_0^\prime} \label{eq:Ssigma_coeffs1} \\
\lambda_n = \frac{n^2-1}{n^2+1},\quad \chi_n = \frac{2\alpha_0^{\prime\prime}}{n\alpha_0^\prime\beta_0^{\prime\prime}}-\frac{8}{n\left(n^2+1\right)\beta_0^\prime}+\frac{2}{n\alpha_0^\prime} \label{eq:Ssigma_coeffs2}
\end{gather}
where the elliptic integrals for the case of an ellipsoid of revolution are
\begin{gather}
\alpha_0 = \int_0^\infty\frac{ds}{\left(n^2+s\right)Q_n\left(s\right)},\quad \beta_0=\int_0^\infty\frac{ds}{\left(1+s\right)Q_n\left(s\right)} \\
\alpha_0^\prime = \int_0^\infty\frac{ds}{\left(1+s\right)^2Q_n\left(s\right)},\quad \beta_0^\prime=\int_0^\infty\frac{ds}{\left(n^2+s\right)\left(1+s\right)Q_n\left(s\right)} \\
\alpha_0^{\prime\prime} = \int_0^\infty\frac{s ds}{\left(1+s\right)^2Q_n\left(s\right)},\quad \beta_0^{\prime\prime}=\int_0^\infty\frac{s ds}{\left(n^2+s\right)\left(1+s\right)Q_n\left(s\right)}
\end{gather}
with $Q_n\left(s\right)=\left(1+s\right)\sqrt{n^2+s}$.

\begin{figure}[tph]
  \setlength{\tabcolsep}{0pt}
  \begin{tabular}{cc}
    \includegraphics[width=0.245\textwidth]{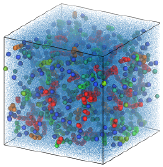}&  
    \includegraphics[width=0.245\textwidth]{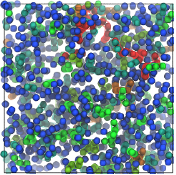}\\
    \textit{(a)}&
    \textit{(b)}\\
    \includegraphics[width=0.245\textwidth]{zero_top_4.pdf}&
    \includegraphics[width=0.245\textwidth]{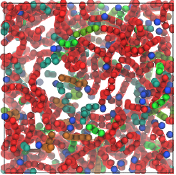}\\
    \textit{(c)}&
    \textit{(d)}
  \end{tabular}
  \caption{\textit{(a)} - simulation snapshot (box size $24\sigma\times 24\sigma\times 24\sigma$) of spontaneous association of dipolar particles without an applied field at a volume fraction $\phi=5\%$ and dipole interaction parameter $\lambda=4$. Blue background depicts a sea of MPCD point particles representing the solvent.
  Simulation snapshots for varying strength of dipolar interactions: \textit{(b)} $\lambda=3$, \textit{(c)} $\lambda=4$, \textit{(d)} $\lambda=5$. Substantial lengthening and branching of the particle chains is observed at larger $\lambda$. Particles color-coded by chain length $n$: monomers ($n=1$) are blue, $n\geq10$ - red.  
}
  \label{fgr:Szero_snapshots}
\end{figure}

\begin{figure}[tph]
\centering
  \setlength{\tabcolsep}{0pt}
  \begin{tabular}{cc}
    \includegraphics[width=0.4\linewidth]{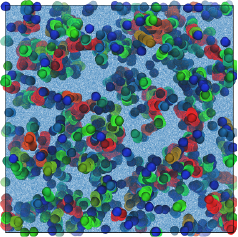}&
    \includegraphics[width=0.4\linewidth]{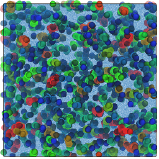}\\
    \textit{(a)}&
    \textit{(b)}\\
    \includegraphics[width=0.4\linewidth]{img_mol0p0.pdf}&
    \includegraphics[width=0.4\linewidth]{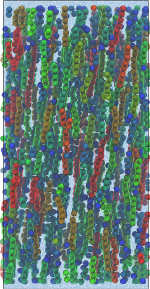}\\
    \textit{(c)}&
    \textit{(d)}
  \end{tabular}
  \caption{Simulation snapshots (box size $24\sigma\times 48\sigma\times 24\sigma$) in strong vertical external field at a volume fraction $\phi=5\%$ and dipole interaction parameter $\lambda=4$ in equilibrium (left - $\left(a\right)$, $\left(c\right)$) and under shear (right - $\left(b\right)$, $\left(d\right)$) show chains of various length aligned in the direction of the field and deviated by imposed shear flow at $Mn\approx0.012$. Side view - $\left(c\right)$, $\left(d\right)$ along the vorticity direction. Top view - $\left(a\right)$, $\left(b\right)$ along the direction of the applied field showing that the chains are uniformly dispersed within the layer in either case. Particles color-coded by chain length $n$: monomers ($n=1$) are blue, $n\geq10$ - red. }
  \label{fgr:Sinf_snapshots}
\end{figure}

\begin{figure}[t]
\centering
  \includegraphics[width=0.7\linewidth]{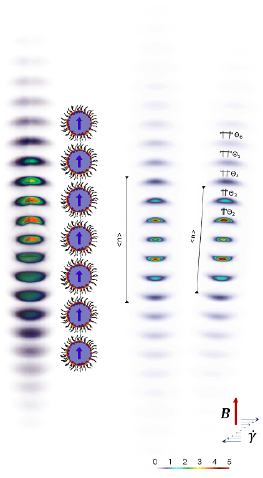}
  \caption{Calculated positional probability density distribution functions $P\left(\bm{r}\right)$ corresponding to the simulations in Figure~\ref{fgr:Sinf_snapshots} ($\lambda=4$, $\phi=5\%$, note: free-floating monomer contributions are substracted). Left - 3D PDF $P\left(\bm{r}\right)$ in equilibrium. Right - projection of the PDF $P\left(\bm{r}\right)$ onto the shear (velocity-vorticity) plane for the same system in equilibrium and at a shear rate $Mn\approx0.012$. The deviation of the $n-$particle chains by the angle $\theta_n$ from the direction of the field is visible on the PDF. The bars show the average chain length $\left\langle n\right\rangle$ for comparison.}
  \label{fgr:Spdf_chain}
\end{figure}

\begin{figure}[t]
\centering
  \includegraphics[width=0.7\linewidth]{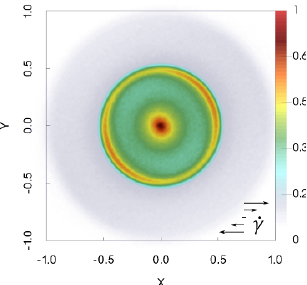}
  \caption{Projection of the non-equilibrium PDF $P\left(\bm{r}\right)$ onto the shear (velocity-vorticity) plane without an applied field at a volume fraction $\phi=5\%$, dipole interaction parameter $\lambda=4$ and $Mn\approx 0.016$, showing the deformation of the conformations ellipsoid in simple shear.  
}
  \label{fgr:Szero_prob}
\end{figure}

\begin{figure}[t]
\centering
    \includegraphics[width=0.8\linewidth]{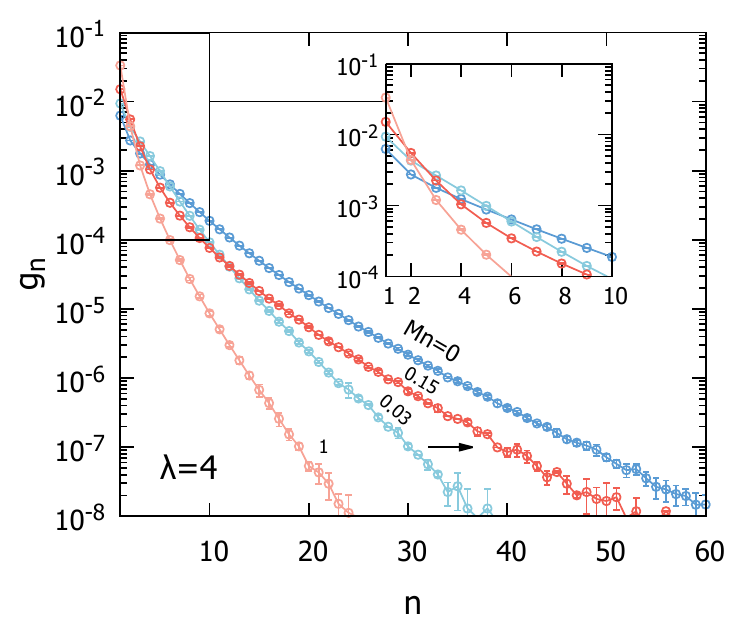}
  \caption{Top: $\left(a\right)$ - chain length distributions $g_n$ in simple shear (with reduced shear rate $Mn$) and strong external field for $\phi=5\%$ and $\lambda=4$. The inset shows a detailed view of the initial region. The recovery of the tail of the chain length distribution is observed at approx. $0.03<Mn<0.15$ signifying a transition into a shear-banded state.}
  \label{fgr:Sg4_Mn}
\end{figure}

\begin{figure}[t]
\centering
  \includegraphics[width=0.6\linewidth]{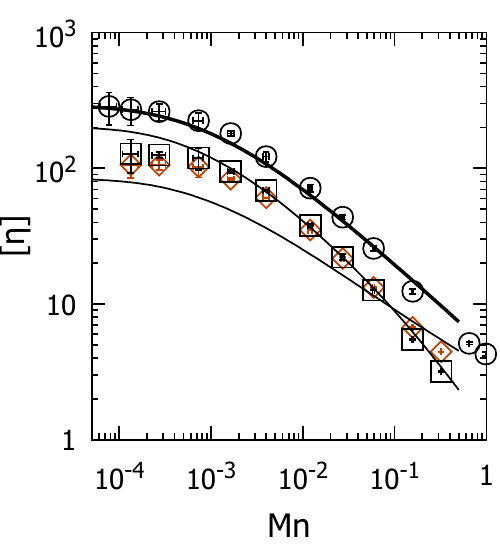}
  \caption{Contributions of the symmetric ($\Box$) and antisymmetric ($\diamond$) stress to the intrinsic viscosity $\left[\eta\right]$ ($\circ$) at $\lambda=4$ ($\phi=5\%$). The potential contribution is overestimated by \eqref{eq:sigma_a}, whereas \eqref{eq:sigma_s} underestimates the hydrodynamic stress. Nevertheless, overall there is good agreement considering the strength of the approximations. Both contributions to the total stress are of similar magnitude at $\lambda=4$. }
  \label{fgr:Svisc_asym}
\end{figure}

\begin{figure}[t]
\centering
  \includegraphics[width=0.6\linewidth]{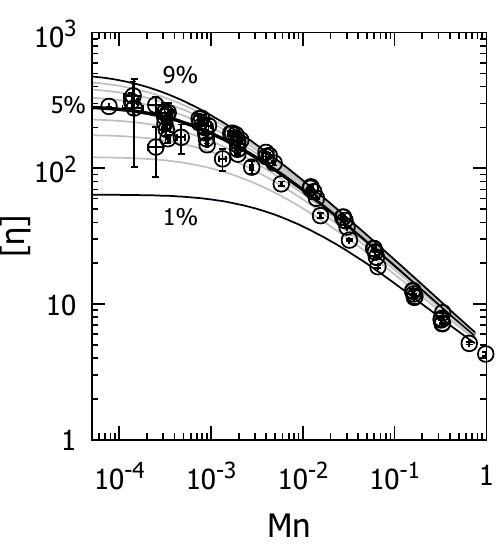}
  \caption{Intrinsic viscosity $\left[\eta\right]$ as a function of the Mason number $Mn$ for varying fraction of dipolar particles $\phi=1\%-9\%$ at $\lambda=4$.  
  A series of calculations has been performed with varying volume fractions $\phi=1\%-9\%$ of the colloidal particles to check whether the intrinsic viscosity \eqref{eq:intrinsic} provides an appropriate scaling for a self-assembled system, where the contributions of the individual colloidal particles strictly speaking should not be additive. The figure shows that for all considered volume fractions the shear thinning regions satisfactorily collapse onto a single curve, which is supported by the chain model. In turn, the model predicts that the plateau region scales as  $\left[\eta\right]\propto \phi^{2}$. The simulation data also shows some dispersion in the vicinity of plateau, however, much smaller than the theoretically predicted. Thus, the intrinsic viscosity provides a satisfactory scaling with respect to the concentration $\phi$ of the colloidal particles along the whole viscometric curve.}
  \label{fgr:Svisc_conc}
\end{figure}

\end{document}